\pdfoutput=1
\documentclass[11pt]{article}
\usepackage[]{acl}
\usepackage{color}
\usepackage{bbm}
\usepackage{multirow}
\usepackage[inline]{enumitem}
\usepackage{graphicx}
\usepackage{subfigure} 
\usepackage{sistyle}
\usepackage[ruled]{algorithm2e}
\usepackage{booktabs}
\usepackage{caption}
\usepackage{graphicx} 
\SIthousandsep{,}
\usepackage{makecell}
\usepackage[skip=0pt]{caption}
\usepackage{amsmath}
\usepackage{hyperref}
\usepackage{array} 
\usepackage{graphicx}
\usepackage{acronym}
\usepackage[export]{adjustbox}
\usepackage{booktabs}
\usepackage{times}
\usepackage{latexsym}

\usepackage[T1]{fontenc}

\usepackage[utf8]{inputenc}

\usepackage{microtype}

\usepackage{inconsolata}

\usepackage{graphicx} 
\usepackage{nccmath}
\usepackage{balance}

\newcommand{\heading}[1]{\vspace*{0.5mm}\noindent\textbf{#1.}}
\AtBeginDocument{%
  \providecommand\BibTeX{{%
    \normalfont B\kern-0.5em{\scshape i\kern-0.25em b}\kern-0.8em\TeX}}}

\makeatletter
\g@addto@macro\normalsize{%
  \abovedisplayskip 2pt plus1pt 
  \belowdisplayskip 2pt plus1pt
  \abovedisplayshortskip  2pt plus1pt%
  \belowdisplayshortskip  1pt plus1pt
}
\makeatother

\acrodef{IR}{infor\-mation retrieval}
\acrodef{LLM}{large language model}
\acrodef{QA}{question answering}
\newcommand{\orfunction}{SumMargLH\xspace}
\newcommand{\andfunction}{JointLH\xspace}
\newcommand{\singlelh}{SingleLH\xspace}

\newcommand{\randfunction}{Rand1LH\xspace}
\newcommand{\UR}{UtilRank\xspace}
\newcommand{\US}{UtilSel\xspace}
\newcommand{\RS}{RelSel\xspace}

\author{Hengran Zhang \textsuperscript{\rm 1,2}\footnotemark[1]\space\space 
Minghao Tang \textsuperscript{\rm 1,2}\footnotemark[1]\space\space
Keping Bi\textsuperscript{\rm 1,2}\footnotemark[2]\space\space 
Jiafeng Guo\textsuperscript{\rm 1,2}\footnotemark[2]\space\space 
Shihao Liu\textsuperscript{\rm 3}  \space\space \\
\textbf{Daiting Shi}\textsuperscript{\rm 3} \space\space 
\textbf{Dawei Yin}\textsuperscript{\rm 3}\space\space 
\textbf{Xueqi Cheng}\textsuperscript{\rm 1,2 }\space\space\\
\textsuperscript{\rm 1}State Key Laboratory of AI Safety, Institute of Computing Technology, \\
Chinese Academy of Sciences \\
\textsuperscript{\rm 2}University of Chinese Academy of Sciences \textsuperscript{\rm 3} Baidu Inc.\\
\{zhanghengran22z, tangminghao25s, bikeping, guojiafeng, cxq\}@ict.ac.cn,  \\
 \{liushihao02, shidaiting01\}@baidu.com, yindawei@acm.org \\
 }

\renewcommand{\thefootnote}{\fnsymbol{footnote}}
\title{Utility-Focused LLM Annotation for Retrieval and Retrieval-Augmented Generation}

\begin{document}
\maketitle

\begin{abstract} 
This paper explores the use of large language models (LLMs) for annotating document utility in training retrieval and retrieval-augmented generation (RAG) systems, aiming to reduce dependence on costly human annotations. We address the gap between retrieval relevance and generative utility by employing LLMs to annotate document utility. To effectively utilize multiple positive samples per query, we introduce a novel loss that maximizes their summed marginal likelihood. 
Using the Qwen-2.5-32B model, we annotate utility on the MS MARCO dataset and conduct retrieval experiments on MS MARCO and BEIR, as well as RAG experiments on MS MARCO QA, NQ, and HotpotQA. 
Our results show that LLM-generated annotations enhance out-of-domain retrieval performance and improve RAG outcomes compared to models trained solely on human annotations or downstream QA metrics. Furthermore, combining LLM annotations with just 20\% of human labels achieves performance comparable to using full human annotations. 
Our study offers a comprehensive approach to utilizing LLM annotations for initializing QA systems on new corpora. Our code and data are available at \url{https://github.com/Trustworthy-Information-Access/Utility-Focused-LLM-Annotation}. 

\end{abstract}
\footnotetext[1]{Contributed equally}
\footnotetext[2]{Corresponding authors}
\renewcommand{\thefootnote}{\arabic{footnote}}
\setcounter{footnote}{0}

\section{Introduction} 
Information retrieval (IR) has long been essential for information seeking, and retrieval-augmented generation (RAG) is increasingly recognized as a key strategy for reducing hallucinations in large language models (LLMs) in the modern landscape of information access \cite{shuster2021retrieval, zamani2022retrieval, ram2023context}. 
Typically, retrieval models rely on human annotations of query-document relevance for training and evaluation. In RAG, the goal shifts towards optimizing the final question answering (QA) performance using results from effective retrievers, with less emphasis on retrieval performance itself. 
Given the high cost of human annotation and the promising potential of LLMs for relevance judgments \cite{rahmani2024llmjudge}, we aim to explore whether LLM-generated annotations can effectively replace human annotations in training models for retrieval and RAG. This is particularly crucial for initializing QA systems based on a reference corpus without annotations. 


There is a gap between the objectives of retrieval and RAG. Retrieval focuses on topical relevance, while RAG requires reference documents to be useful for generation (i.e., utility).  
In other words, results considered relevant by a retriever may not be useful for an LLM during generation. 
Aware of this mismatch, researchers have shifted from using relevance annotations as document labels to assessing LLM performance on downstream tasks with the document as its label~\cite{shi2024replug, lewis2020retrieval, izacard2023atlas, glass2022re2g, zamani2024stochastic, gao2024smartrag}. This includes metrics such as the likelihood of generating ground-truth answers~\cite{shi2024replug} or exact match scores between generated and ground-truth answers~\cite{zamani2024stochastic}. 
Another approach involves prompting LLMs to select documents with utility from relevance-oriented retrieval results for use in RAG \cite{zhang2024iterative,zhang2024large}. Studies from both approaches have demonstrated improved RAG performance.


Despite their effectiveness, both approaches have limitations. 
The first approach requires manually labeled ground-truth answers to assess downstream task performance, which results in substantial QA annotation costs. Additionally, retrievers trained on the performance of a specific task may struggle to generalize to other downstream tasks or even different evaluation metrics within the same task. This issue is exacerbated when dealing with non-factoid questions, where accurate evaluation is challenging, making it less feasible to use QA performance as training objectives for retrieval. 
In contrast, the second approach, which leverages LLMs to select useful documents for generation~\cite{zhang2024iterative,zhang2024large}, does not require human annotation and is not confined to specific tasks or metrics. However, the selection is from initially retrieved results and cannot scale to the entire corpus during inference due to prohibitive costs.

To address these limitations, this paper proposes using LLMs to annotate document utility for retriever training, aiming to identify useful documents from the entire collection for RAG. 
We focus on four research questions (RQs): (\textbf{RQ1}) What is the optimal training strategy when multiple annotated positive samples are available for a query, in terms of data ingestion and retriever optimization? (\textbf{RQ2}) How do retrievers trained with LLM-annotated utility compare to those trained with human-annotated relevance in both in-domain and out-of-domain retrieval? (\textbf{RQ3}) Can LLM-annotated data enhance retrieval performance when human labels are already available? (\textbf{RQ4}) Do retrievers trained with utility-focused LLM annotations result in better RAG performance compared to those trained with downstream task performance metrics and human annotations in both in-domain and out-of-domain collections?

To study the research questions, we employ a state-of-the-art open-source LLM, Qwen-2.5-32B-Int8 \cite{yang2024qwen2}, to annotate the utility of hard negatives in the MS MARCO dataset \cite{nguyen2016ms}. 
In contrast to human annotation on MS MARCO, which has one positive sample per query, Qwen annotates an average of 2.9 positive samples per query. Optimizing the standard joint likelihood of the multiple positives results in significant performance regression. To address the challenges posed by multiple positives, we introduce a novel loss function, \orfunction, which maximizes their summed marginal likelihood and performs significantly better. 
For retrieval evaluation, we compare retrievers trained with LLM and human annotations on the MS MARCO Dev set and BEIR \cite{thakur2beir}. For RAG evaluation, we assess the retrievers on the MS MARCO QA task and two QA tasks with retrieval collections also included in BEIR, i.e., NQ \cite{kwiatkowski2019natural} and HotpotQA \cite{yang2018hotpotqa}. 
Our findings include: 1) LLM annotations alone result in worse in-domain retrieval performance but better out-of-domain performance compared to human annotations; 2) Combining LLM annotations with 20\% of human annotations achieves similar performance to models trained with 100\% human labels; 3) Retrievers trained with both LLM and human annotations using curriculum learning significantly outperform those using only human annotations; 4) The findings for RAG performance are consistent with the retrieval performance regarding both in-domain and out-of-domain datasets. 
We summarize our contributions as follows: 
\begin{itemize}[leftmargin=*,itemsep=0pt,topsep=0pt,parsep=0pt]
\item We introduce a comprehensive solution for data annotation using LLMs for retrieval and RAG, along with corresponding training strategies.
\item We conduct an extensive study on the use of LLM-annotated utility to train retrievers for both in-domain and out-of-domain retrieval and RAG.
\item Extensive experiments and analyses demonstrate the advantages of leveraging utility-focused LLM annotations for retrieval and RAG, particularly for out-of-domain data.
\item We enhance the MS MARCO dataset with LLM annotations, providing passage labels for approximately 500K queries, which can facilitate research on false negatives, weak supervision, and retrieval evaluation by LLMs. 
\end{itemize}
Our work offers a viable and promising solution for initiating QA systems on new corpora, especially when human annotations are unavailable and budgets are limited.

\begin{figure*}[t]
    \centering
    \small
    \includegraphics[width=0.9\linewidth]{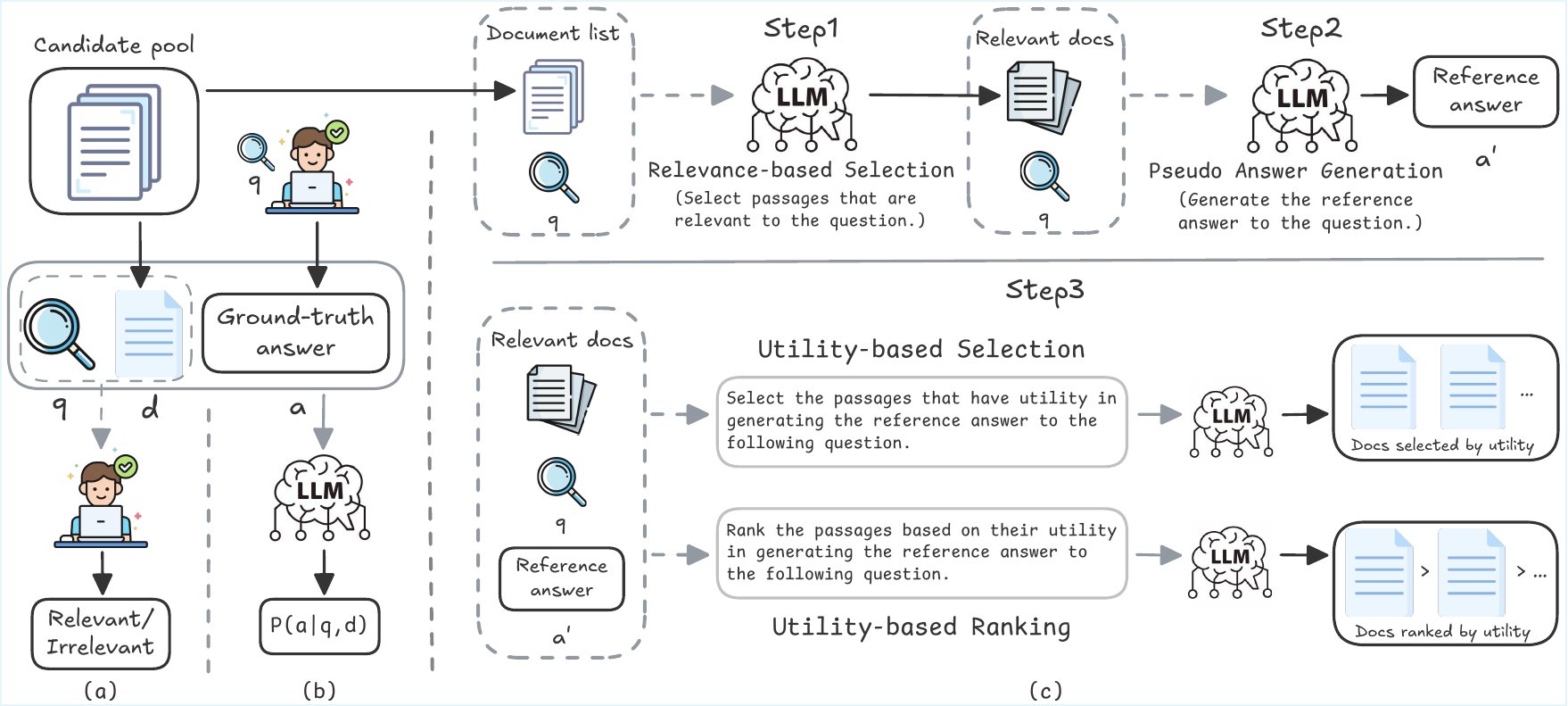}
    \caption{Different annotation methodologies: (a) Human annotation, (b) Using downstream task performance as utility score, (c) Our utility-focused annotation pipeline. The prompts are illustrative, see Appendix \ref{app:prompts} for details.}
    \label{fig:prompt} 
\end{figure*}

\vspace{-1mm}
\section{Related Work}
\vspace{-2mm}
\subsection{First-Stage Retrieval}
Initially, the first-stage retrieval models were predominantly classical term-based models, such as BM25 \cite{robertson2009probabilistic}, which combines term matching with TF-IDF weighting. 
To address the semantic mismatch limitations of classical term-based models, neural information retrieval (IR) emerged by leveraging neural networks to learn semantic representations \cite{huang2013learning, guo2016deep}. 
Subsequently, pre-trained language model (PLM)-based retrievers have been extensively explored \cite{xiao2022retromae, wang2023simlm, izacard2021unsupervised, ma2021b, ren2021rocketqav2}.
More recently, LLMs have been directly applied as first-stage retrieval models \cite{ma2024fine, springer2024repetition,  zhang2025unleashing, li2024llama2vec}, demonstrating unprecedented potential in IR. 


\subsection{Utility-Focused RAG} 
There is a gap between the objectives of retrieval and RAG.
Retrieval focuses on topical relevance, while RAG requires reference documents to be useful for effective generation. 
To address this issue, current research mainly focuses on two approaches: 
\begin{enumerate*}[leftmargin=*,itemsep=0pt,topsep=0pt,parsep=0pt]
    \item 
    Verbalized utility judgments, which directly utilized LLMs for selecting useful documents from the retrieved document list \cite{zhang2024large, zhang2024iterative, zhao2024longrag}. 
    \item Utility-optimized retriever, which involves transferring the preference of LLMs to the retriever. 
    Two primary optimization signals are commonly employed: 
    \begin{enumerate*}[leftmargin=*,itemsep=0pt,topsep=0pt,parsep=0pt]
        \item the likelihood of generating the ground truth answers given the query and document \cite{shi2024replug, lewis2020retrieval, izacard2023atlas, glass2022re2g,  bacciu2023rraml}; 
        \item evaluation metrics of the downstream tasks \cite{zamani2024stochastic, gao2024smartrag, wang2024retrieve}, such as exact match.
    \end{enumerate*}
    This approach relies on ground truth answers for specific downstream tasks and limits generalization. 
\end{enumerate*}

\vspace{-1mm}
\subsection{Automatic Annotation with LLMs}
In the field of information retrieval, many studies \cite{thomas2024large, rahmani2024llmjudge, takehi2024llm, ni2024diras, zhang2024iterative} have explored the annotation capabilities of LLMs for relevance judgments. 
However, these studies predominantly focus on small evaluation datasets, lacking a comprehensive investigation into the annotation capabilities of LLMs to scale to the entire training datasets for retrieval-related task.


\section{Utility-Focused LLM Annotation}
Figure \ref{fig:prompt}(a)\&(b) illustrates two primary types of document labels used in retriever training for RAG: human-annotated relevance labels and utility scores derived from downstream tasks. Retrievers trained using human-annotated relevance typically focus on aboutness and topic-relatedness. In contrast, utility scores, which are estimated based on downstream tasks, such as the probability of LLMs generating the correct answer given a document, are more beneficial for RAG \cite{shi2024replug}. Building on the insight that LLMs can effectively assess utility for RAG \cite{zhang2024large}, we introduce a utility-focused LLM annotation pipeline for training retrievers, as depicted in Figure \ref{fig:prompt}(c). This approach is designed for both initial retrieval stages and RAG, aiming to minimize the manual effort required for annotating document relevance and ground-truth answers.

\subsection{Annotation Methodology} 
\label{subsec:anno-methodology}
\heading{Annotation Pool Construction}
Given a query, the majority of documents in a corpus are irrelevant, making it impractical to annotate the utility of every document with LLMs. A common practice is to compile a candidate pool by aggregating documents retrieved by effective retrievers, such as unsupervised methods like BM25 \cite{robertson2009probabilistic}, and retrievers trained on other collections. 
We adopt a similar approach in our study. Our annotation process is based on the widely used retrieval benchmark, the MS MARCO passage set \cite{nguyen2016ms}. It is well-known that MS MARCO typically includes only one annotated positive example per query and many false negatives due to under-annotation \cite{craswell2020overview,craswell2021overview}. 

Retrievers trained with MS MARCO typically gather a pool of hard negatives $\{d_{i}^{-}\}_{i=1}^{n}$, from which a subset of $m$ samples is randomly selected. These sampled hard negatives, along with the single positive $d^{+}$ and in-batch negatives, are then used for contrastive learning. 
To neutralize the impact of hard negatives when comparing the retrievers trained with human and LLM annotations, we utilize the same collection of positives and hard negatives as in \citet{ma2024fine} (from BM25 and CoCondenser \cite{gao2021rethink}) for LLM annotation. This ensures that all comparison models have the same set of $n+1$ annotated documents for each query, differing only in their annotations. $m+1$ instances are selected for training in each epoch, including positives and randomly sampled negatives ($n=30,m=15$ in this paper). 
To study the effect of whether human-annotated positives are included in the annotation pool, we compare the performance of consistently including and excluding human-annotated positives in training. As presented in Appendix \ref{app:human_must_yes_no}, the essential conclusions are similar to those we report in Section \ref{sec:exp_results}.


\begin{figure}[t]
    \centering
    \includegraphics[width=\linewidth]{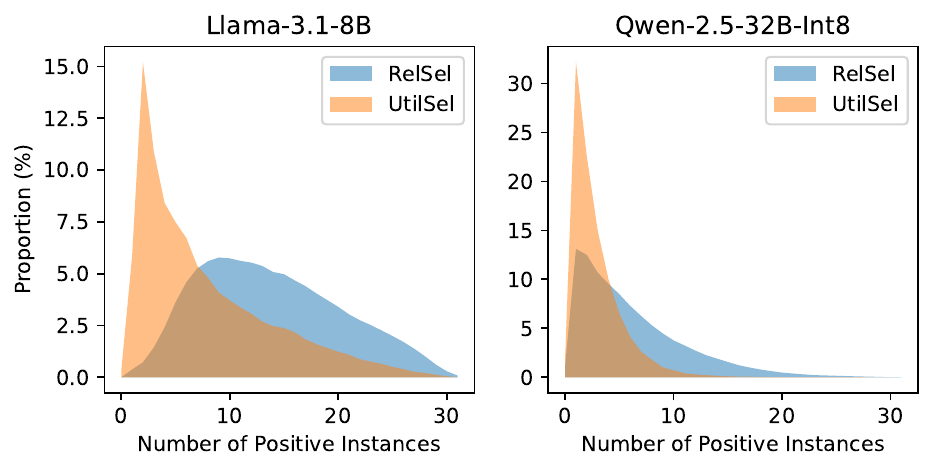}
    \caption{Positive annotation distribution of different annotators at various stages.}
    \label{fig:freq_distribution}
\end{figure}

\heading{Annotation Methods} 
After collecting the candidate pool, we apply three annotation methods, as illustrated in Figure \ref{fig:prompt}(c): relevance-based selection (\RS), utility-based selection (\US), and utility-based ranking (\UR). 
In \textbf{\RS}, we begin with an initial filtering step where an LLM is used to select a subset of documents that are topically relevant to the query. Next, we employ the utility judgment method from \citet{zhang2024large}, which involves generating a pseudo-answer based on the output from \RS and assessing document utility for downstream generation using the pseudo-relevant documents and pseudo-answer. This listwise comparison enables the LLM to make accurate relative judgments. In \textbf{\US}, the LLM selects the subset of useful documents. In contrast, \textbf{\UR} asks the LLM to rank the input documents according to their utility, then the top $k\%$ documents are annotated as positive ($k=10$ in our main experiments). The float number is rounded down, and if the result is zero, a single document will be marked as positive. 
\US can flexibly determine the number of useful documents, whereas \UR allows for different thresholds to balance the precision and recall of LLM annotations. All the annotation prompts are detailed in Appendix \ref{app:prompts}.

\subsection{Statistics of LLM Annotations}
\label{subsec:stats_llm_ann}
We employ two well-known open-source LLMs of different sizes for annotation: LlaMa-3.1-8B-Instruct (Llama-3.1-8B) \cite{dubey2024llama} and Qwen-2.5-32B-Instruct with GPTQ-quantized \cite{frantar2022gptq} 8-bit version (Qwen-2.5-32B-Int8) \cite{yang2024qwen2}.

\heading{Positive Annotation Distribution} 
Figure \ref{fig:freq_distribution} shows the distribution of positive annotations made by \RS and \US (\UR is not shown since its number of positives is determined by the threshold $k\%$). The average number section in Table \ref{tab:annotationRecall} provides the specific average number of positive annotations. We find that the instances considered useful by LLMs are significantly fewer than those they identify as relevant, consistent with the findings in \citet{zhang2024iterative}. Additionally, the stronger model (i.e., Qwen) tends to select fewer useful documents.


\heading{Annotation Quality Evaluation}
We compare the consistency of annotations by LLMs and humans. Considering human labels as the ground-truth, the precision and recall of the LLM-marked positives for each method are shown in Table \ref{tab:annotationRecall}. It reveals that 1) \US has higher precision and lower recall than \RS, 2) Qwen is more accurate than Llama in selecting the human positive (precision doubled with some real drop). 
As we know, there are false negatives in the annotation pool. We also manually checked around 200 LLM annotations and found that LLM-annotated positives are more than actual positives. This means that LLM should be stricter to be more accurate. Qwen has fewer false-positive issues, and its \UR has the best overall precision and recall trade-off. 
Since Qwen has better annotation quality, our experiments in Section \ref{sec:exp_results} are all based on its annotations.

\begin{table}[t]
  \centering
  \small
    \renewcommand{\arraystretch}{0.9}
   \setlength\tabcolsep{2pt}
    \begin{tabular}{l ccc ccc ccc}
    \toprule
    \multicolumn{1}{c}{\multirow{2}[4]{*}{LLM}} & \multicolumn{3}{c}{Precision} & \multicolumn{3}{c}{Recall} & \multicolumn{3}{c}{Avg Number}\\
\cmidrule(r){2-4} \cmidrule(r){5-7}   \cmidrule(r){8-10}         & \multicolumn{1}{l}{RS} & \multicolumn{1}{l}{US} & \multicolumn{1}{l}{UR} & \multicolumn{1}{l}{RS} & \multicolumn{1}{l}{US} & \multicolumn{1}{l}{UR} & \multicolumn{1}{l}{RS} & \multicolumn{1}{l}{US} & \multicolumn{1}{l}{UR} \\
    \midrule
       Llama  & \phantom{1}7.1 & 11.9 &    36.5 &  97.6  &  91.6  & 41.0     & 13.8 & 7.7 &  1.2\\
       Qwen &  15.1  &  29.5  &  71.3  &  92.8  &  84.8  &  72.0     & \phantom{1}6.2  & 2.9 & 1.0 \\
    \bottomrule
    \end{tabular}
    \caption{Precision and Recall (\%) of human positive under different annotations. ``RS'', ``US'', ``UR'' mean ``\RS'', ``\US'', ``\UR'', respectively. }
  \label{tab:annotationRecall}
\end{table}

\subsection{Training with Utility Annotations}
\heading{Loss Function}
Dense retrievers are typically trained to maximize the likelihood of a positive sample $d+$ compared to a negative passage set $D^-$, which usually includes hard negatives and in-batch negatives \cite{karpukhin2020dense}. Given a query $q$, the probability of a document $d$ to be positive in $\{d^+\} \cup D^-$ is calculated as: 

\begin{equation}
	\small
	P(d|q, d^+, D^-) = \frac{\exp(s(q, d))} {\sum_{d' \in { \{d^+\}\cup D^-} } \exp(s(q, d'))}, 
\end{equation}
where $s(q,d)$ is the matching score of $q$ and $d$. 

\textsf{\singlelh}. As many large-scale retrieval datasets, such as MS MARCO, only have one relevant instance per query, the loss function is usually maximizing the likelihood of the single positive:

\begin{equation}
\small
\mathcal{L}_{s}(q, d^+, D^-) = - \log P(d^+|q, d^+, D^-).
\end{equation}
Since LLMs have multiple positive annotations, \singlelh cannot be used directly. 

\textsf{\randfunction}. A straightforward approach is to randomly sample one positive instance per query in each epoch and use the standard \singlelh for training, which we name as \randfunction. 

\textsf{\andfunction}. Another common way is to enlarge $\{d^+\}$ to a positive passage set $D^+  (|D^+| \geq 1)$ and optimize the joint likelihood of each positive instance in $D^+$:
\begin{equation}
	\small
	\mathcal{L}_{s}(q, D^+, D^-) = - \log \prod_{d^+ \in D^+} P(d^+|q, D^+, D^-).
	\label{eq:joint-loss}
\end{equation}
This function may not be robust to low-quality annotations, as even a single false positive can significantly affect the overall loss. As noted in Section \ref{subsec:stats_llm_ann}, LLM annotations include false positives, which can make this loss function suboptimal.

\textsf{\orfunction}. Considering the quality of LLM annotation may be unstable, we propose a novel objective that maximizes the summed marginal likelihood of each positive instance in $D^+$, i.e.,  

\begin{equation}
	\small
	\mathcal{L}_{s}(q, D^+, D^-) = - \log \sum_{d^+ \in D^+} P(d^+|q, D^+, D^-).
	\label{eq:sum-margin-loss}
\end{equation}
It optimizes the overall likelihood of instances in $D^+$ to be positive, and does not require the likelihood of each positive to be maximized. Thus, it relaxes the optimization towards potentially false positives, and can better leverage LLM annotations (shown in Section \ref{sec:further_analysis}).

\heading{Combining Human and LLM Annotations}
\label{heading:curriculum learning}
When budgets allow, human-labeled data can be used alongside LLM annotations rather than relying solely on the latter. Given that human annotations typically have higher quality than those from LLMs, simply merging and treating them equally may not be effective. Therefore, we propose using \textit{curriculum learning} \cite{bengio2009curriculum} (\textbf{CL}) to integrate the two types of data, starting with training retrievers on the lower-quality LLM annotations and subsequently refining them with higher-quality human annotations.

\vspace{-1mm}
\section{Experimental Setup}

\vspace{-1mm}
\subsection{Datasets}
\heading{Retrieval Datasets} 
As in many existing works \cite{xiao2022retromae,guo2022semantic}, we train all retrievers on the MS MARCO training set, with about 503K queries and 8.8 million passages. Retrieval evaluation is conducted on the MS MARCO Dev set, TREC DL 19/20 \cite{craswell2020overview,craswell2021overview} with more human annotations, and the 14 public retrieval datasets across various domains with diverse downstream tasks in BEIR \cite{thakur2beir} benchmark, excluding MS MARCO. 

\heading{RAG Datasets}
We use the MS MARCO QA, which has the ground-truth answers for the queries in the MS MARCO retrieval dataset, to evaluate the RAG performance when using Llama-3.1-8B and Qwen-2.5-32B-Int8 as generators. Similarly, for two subsets of BEIR, i.e., NQ \cite{kwiatkowski2019natural} and HotpotQA \cite{yang2018hotpotqa}, we use the ground-truth answers of the questions to evaluate the RAG performance with the two generators. 
Detailed information about the datasets can be found in Appendix \ref{app:dataset}.

\begin{table*}[t]
\centering
\small
\setlength\tabcolsep{2.5pt}
\begin{tabular}{lllllllllllll}
    \toprule
    \multicolumn{1}{l}{\multirow{4}[8]{*}{Annotation}} & \multicolumn{6}{c}{RetroMAE}        & \multicolumn{6}{c}{Contriever} \\
    
    \cmidrule{2-13}  \multicolumn{1}{c}{} & \multicolumn{4}{c}{Human Test} & \multicolumn{2}{c}{Hybrid Test} & \multicolumn{4}{c}{Human Test} & \multicolumn{2}{c}{Hybrid Test} \\
    
    \cmidrule(r){2-5} \cmidrule(r){6-7}  \cmidrule(r){8-11} \cmidrule(r){12-13}    \multicolumn{1}{c}{} & \multicolumn{2}{c}{Dev} & \multicolumn{1}{c}{DL19} & \multicolumn{1}{c}{DL20} & \multicolumn{1}{c}{\multirow{2}[4]{*}{M@10}} & \multicolumn{1}{c}{\multirow{2}[4]{*}{N@10}} & \multicolumn{2}{c}{Dev} & \multicolumn{1}{c}{DL19} & \multicolumn{1}{c}{DL20} & \multicolumn{1}{c}{\multirow{2}[4]{*}{M@10}} & \multicolumn{1}{c}{\multirow{2}[4]{*}{N@10}} \\
    
    \cmidrule(r){2-3}  \cmidrule(r){4-5} \cmidrule(r){8-9} \cmidrule(r){10-11}   \multicolumn{1}{c}{} & \multicolumn{1}{c}{M@10} & \multicolumn{1}{c}{R@1000} & \multicolumn{1}{c}{N@10} & \multicolumn{1}{c}{N@10} &       &       & \multicolumn{1}{c}{M@10} & \multicolumn{1}{c}{R@1000} & \multicolumn{1}{c}{N@10} & \multicolumn{1}{c}{N@10} &       &  \\
    \midrule
    
    Human & 38.6 & 98.6 & 68.2 & \textbf{71.6} & 83.7  & 63.1  & 35.6 & 97.6 & 68.5  & 67.9  &  82.2  & 62.0 \\
    \midrule
    
    REPLUG & 33.8$^{-}$   & 94.7$^{-}$  & 65.5  & 58.7  & 75.7$^{-}$  & 54.3$^{-}$  &  31.4$^{-}$   &  93.1$^{-}$  &  64.3  &   59.7  &   79.4  &  53.2$^{-}$ \\
    UtilSel & 35.3$^{- \dagger}$  & 97.7$^{- \dagger}$  & \underline{68.0}    & \underline{71.0}    & \textbf{\underline{87.5}}$^{+ \dagger}$ & 65.8$^{+ \dagger}$  & 33.3$^{- \dagger}$  & 96.8$^{- \dagger}$  & 67.8  & 67.8 &  \underline{85.0}$^{\dagger}$  &  63.7$^{\dagger}$\\
    UtilRank & \underline{35.7}$^{- \dagger}$  & \underline{97.8}$^{- \dagger}$  & 67.1  & 71.0    & 86.1$^{\dagger}$  & \textbf{\underline{66.1}}$^{+ \dagger}$ & \underline{33.6}$^{- \dagger}$  & \underline{96.8}$^{- \dagger}$  & \textbf{\underline{70.8}} & \underline{68.8}  &  84.6$^{\dagger}$  & \underline{63.7}$^{\dagger}$ \\
    \midrule
    
    REPLUG (CL 20\%) & 36.6$^{-}$  & 98.3$^{-}$  & 69.5  & 67.8  & 81.7  & 60.2$^{-}$  & 33.7$^{-}$  & 97.2$^{-}$  & 68.4 & 66.6  &  82.9  & 59.4$^{-}$ \\
    UtilSel (CL 20\%) & 38.2$^{\dagger}$  & \underline{98.5}$^{\dagger}$ & 69.6  & \underline{71.4} & 83.4  & \underline{65.5}$^{+ \dagger}$ & 35.3$^{\dagger}$  & 97.4 & 69.3  & 68.7  &  85.4$^{+}$  &  63.4$^{\dagger}$\\
    UtilRank (CL 20\%) & \underline{38.3}$^{\dagger}$ & 98.4  & \textbf{\underline{70.5}} & 70.0    & \underline{84.3} & 64.6$^{\dagger}$  & \underline{35.6}$^{\dagger}$ & \underline{97.4}  & \underline{70.4 } & \textbf{\underline{70.1}} &  \textbf{\underline{86.1}}$^{+}$  & \textbf{\underline{64.0}}$^{\dagger}$\\
    \midrule
  
    REPLUG (CL 100\%) & 38.7  & 98.6  & 69.5  & 69.7  & 83.7  & 63.1  & 35.5  & 97.7  & 68.0  & 69.1  &   80.7   &  59.0$^{-}$\\
    UtilSel (CL 100\%) & \textbf{\underline{39.3}}$^{+ \dagger}$ & 98.6  & \textbf{\underline{70.5}} & \underline{70.9} & \underline{84.7} & \underline{64.7}$^{+ \dagger}$ & \textbf{\underline{36.6}}$^{+ \dagger}$ & \textbf{\underline{97.8}} & 69.3  & 68.4  &  \underline{85.7}$^{+ \dagger}$  & 63.8$^{+ \dagger}$ \\
    UtilRank (CL 100\%) & 39.2$^{+ \dagger}$  & \textbf{\underline{98.7}} & 69.6  & 69.9  & 84.2  & 64.2  & 36.5$^{+ \dagger}$ & \textbf{\underline{97.8}}  & \underline{69.9} & \underline{69.2} &  85.2$^{+ \dagger}$ & \underline{63.9}$^{+ \dagger}$  \\
    \midrule
    
\end{tabular}
\caption{Retrieval performance (\%) of different annotation methods. ``M@k'', ``R@k'', ``N@k'' mean ``MRR@k'', ``Recall@k'', and ``NDCG@k'' respectively. ``$^{+}$'', ``$^{-}$'', and ``$^{\dagger}$'' indicate significant improvements and decrements over Human, and significant improvements over REPLUG within the same group, respectively, using a two-sided paired t-test ($p<0.05$). \underline{underline} and \textbf{Bold} indicate the best performance within each group and overall.}
\label{tab:diff_annotation}
\end{table*}

\vspace{-1mm}
\subsection{Baselines}
Our comparisons of data annotation methods are based on the pretrained version of two representative retrievers, RetroMAE \cite{xiao2022retromae} and Contriever \cite{izacard2021unsupervised} (before fine-tuning). Our baselines include retrievers trained with human annotations and downstream task performance (shown in Figure \ref{fig:prompt}(a)\&(b) respectively):

\begin{itemize}[leftmargin=*,itemsep=0pt,topsep=0pt,parsep=0pt]
    \item \textbf{Human}: Retrievers trained with original human annotations in MS MARCO using \singlelh. 
    \item \textbf{REPLUG} \cite{shi2024replug}: The likelihood of the ground-truth answer given each passage is used as its utility label. Retrievers are optimized towards negative KL divergence between the distribution of passage utility labels and their relevance scores (see Appendix~\ref{app:ds-utility} for details). 
    
    \item \textbf{REPLUG (CL 20\%/100\%)}: This approach initially trains the model with utility scores and then updates the model with either 20\% randomly selected or 100\% of the human annotations using curriculum learning.
\end{itemize}
Similarly, our methods include using LLM annotations alone (\US, \UR), and combining them with 20\%/100\% human annotations using curriculum learning. 
Implementation details of each method can be found in Appendix~\ref{app:implementation}.

\begin{table*}[t]
\centering
\small
\setlength\tabcolsep{2.5pt}
\begin{tabular}{cccccc cccccc}
    \toprule
    \multicolumn{1}{c}{\multirow{2}[4]{*}{Datasets}} & \multicolumn{1}{c}{\multirow{2}[4]{*}{BM25}} & \multicolumn{1}{c}{\multirow{2}[4]{*}{Human}} & \multicolumn{1}{c}{\multirow{2}[4]{*}{REPLUG}}  & \multicolumn{1}{c}{\multirow{2}[4]{*}{\US}} & \multicolumn{1}{c}{\multirow{2}[4]{*}{\UR}} & \multicolumn{3}{c}{Curriculum Learning, 20\%} & \multicolumn{3}{c}{Curriculum Learning, 100\%} \\
    \cmidrule(r){7-9}   \cmidrule(r){10-12}            &       &       &       &       & \multicolumn{1}{c}{} & \multicolumn{1}{c}{REPLUG} & \multicolumn{1}{c}{\US} & \multicolumn{1}{c}{\UR} & \multicolumn{1}{c}{REPLUG} & \multicolumn{1}{c}{\US} & \multicolumn{1}{c}{\UR} \\
    
    \midrule
    DBPedia & 31.8  & 36.0   & 29.1    & \textbf{38.0}   & \underline{37.9}  & 35.9  &  37.4   & 37.4   & 36.1   & 37.1   & 37.5  \\
    FiQA  & 23.6  & 29.7  & 24.9  & \textbf{32.6}  & 31.6  & 30.8 & \underline{32.1}  & 31.3  & 31.3 & 31.6  & 30.4  \\
    NQ    & 30.6  & 49.2  & 41.2   & \underline{53.5} & \textbf{53.9} & 48.0  & 51.4  & 51.9  & 50.1  &  51.9  & 51.7  \\
    HotpotQA & \textbf{63.3} & 58.4  & 57.4   & 59.6  & 59.6  & 60.2  & 60.0  & 59.8  & \underline{60.5}  & 60.1  & 59.5  \\
    NFCorpus & 32.2  & 32.8  & 30.3    & 33.9  & \underline{34.0} & 33.9  & \textbf{34.2}  & 33.8  & 33.7 & \underline{34.0}  & 33.4  \\
    T-COVID & 59.5  & 63.4  & 54.2   & 66.1  & 64.5  & \underline{68.5}  & 65.0  & 67.5  & \textbf{71.8}  & 64.8  & 68.0 \\
    Touche & \textbf{44.2} & 24.2  & 18.9   & \underline{28.5} & 26.6 & 27.0  & 24.7  & 28.0  & 25.4  & 22.6  & 25.7  \\
     CQA   & 32.5  & 32.2  & 29.2   & 32.3  & 30.7 & \underline{33.2}  & \textbf{33.9} & 33.0  & 32.8  & 32.9  & 32.8  \\
    ArguAna & \textbf{39.7} & 30.5  & 22.7  & 34.1  & 25.0  & 32.9  & \underline{36.4}  & 29.3  & 29.0  & 30.8  & 28.1  \\
    C-FEVER & 16.5  & 18.0  & 13.2   & \textbf{19.5} & 16.4  & 17.9  & 16.5  & 15.3  & 18.4  & \underline{18.5}  & 16.8  \\
    FEVER & 65.1  & 66.6  & 66.1  & \textbf{73.8}  & \underline{73.1}  & 72.3  & 69.9  & 72.4  & 71.1  & 70.1  & 71.0  \\
    Quora & 78.9  & 86.2  & 76.9    & 85.4 & 85.3  & 85.3  & 86.1  & 85.9  & 85.7  & \underline{86.4}  & \textbf{86.5 } \\
    SCIDOCS & 14.1 & 13.4  & 13.5  & 14.3  & 13.6  & \textbf{14.5}  & \underline{14.4}  & 13.9  & 13.9  & 13.7  & 13.6  \\
     SciFact & \textbf{67.9 } & 63.1  & 59.3 & 62.8 & 63.2 & 63.2  & 64.2  & 63.8  & 63.6  & 64.1  & \underline{64.9}  \\
    \midrule
     Average  & 42.9  & 43.1 & 38.4  & \textbf{45.3} & 43.9  & 44.5  & \underline{44.7}  & 44.5 & 44.5  & 44.2 & 44.3  \\
    
    \bottomrule
\end{tabular}
\caption{Zero-shot retrieval performance (NDCG@10, \%) of different retrievers (RetroMAE backbone) trained with various annotations. \textbf{Bold} and \underline{underlined} represent the best and second best performance, respectively.}
\label{tab:beir_zero_shot}
\end{table*}

\subsection{Evaluation} 
Human annotations often contain many false negatives due to under-annotation, and humans may have different preferences from LLMs. Evaluating retrieval performance using human labels as ground truth may be unfair to models trained with LLM annotations. To create a more balanced comparison set with more relevance labels and fewer false negatives, we randomly sampled 200 queries from the MS MARCO Dev set. For each query, we collected a candidate pool by merging the top 20 retrieved passages from various retrievers (Human, REPLUG, \US, \UR) and used GPT-4o-mini \cite{hurst2024gpt} to select positive instances from the pool based on the \textit{ground-truth} answer, using the \US prompt (see Appendix~\ref{app:prompts}). Both the original human and GPT-annotated positives are considered new golden labels. We refer to this combined set as the \textbf{Hybrid Test} and the set with only human annotations as the \textbf{Human Test}.

We evaluate retrievers trained with MS MARCO annotated data by humans or LLMs under both in-domain settings (MS MARCO Dev, TREC DL 19/20, MS MARCO Hybrid Test) and out-of-domain settings (14 BEIR datasets). The retrieved results are then directly fed to generators to assess downstream QA performance on MS MARCO QA and two BEIR datasets, NQ and HotpotQA. Detailed evaluation metrics for retrieval and RAG are provided in Appendix~\ref{app:evaluation}.

\begin{table*}[t]
    \centering
    \small
    \setlength\tabcolsep{2pt}
    \begin{tabular}{ l l llll llll}
    \toprule
    
    \multicolumn{1}{l}{\multirow{2}[3]{*}{Annotation}} & \multicolumn{1}{l}{\multirow{2}[3]{*}{Recall}} & \multicolumn{4}{c}{Generator: Llama-3.1-8B} & \multicolumn{4}{c}{Generator: Qwen-2.5-32B-Int8} \\
    \cmidrule(r){3-6} \cmidrule(r){7-10}    \multicolumn{1}{c}{}
    &       & \multicolumn{1}{l}{BLEU-3} & \multicolumn{1}{l}{BLEU-4} & \multicolumn{1}{l}{ROUGE-L} & \multicolumn{1}{l}{BERT-score} & \multicolumn{1}{l}{BLEU-3} & \multicolumn{1}{l}{BLEU-4} & \multicolumn{1}{l}{ROUGE-L} & \multicolumn{1}{l}{BERT-score} \\
    
    \midrule
    Human & 24.7  & 17.2  & 14.2  & 35.7  & 67.8  & 15.8  & 12.6  & 34.3  & 67.4  \\
    \midrule
    REPLUG & 21.7$^{-}$  & 15.7  & 12.9  & 33.8$^{-}$  & 66.7$^{-}$  & 14.7  & 11.6  & 32.4$^{-}$  & 66.2$^{-}$  \\
    \US & 22.3$^{-}$  & 16.3  & 13.4  & 34.7$^{- \dagger}$  & 67.4$^{- \dagger}$  & 14.9  & 11.7  & 33.5$^{- \dagger}$  & 67.1$^{- \dagger}$  \\
    \UR & \underline{22.6}$^{-}$  & \underline{16.6}  & \underline{13.6}  & \underline{35.1}$^{- \dagger}$  & \underline{67.5}$^{- \dagger}$  & \underline{15.2}  & \underline{12.0}  & \underline{33.9}$^{- \dagger}$  & \underline{67.3}$^{- \dagger}$  \\
   
    \midrule
    REPLUG (CL 20\%) & 23.2$^{-}$  & 16.7  & 13.7  & 34.9$^{-}$  & 67.4$^{-}$  & 15.2  & 12.1  & 33.6$^{-}$  & 67.1$^{-}$  \\ 
    \US (CL 20\%) & \underline{24.6}$^{\dagger}$  & \underline{17.4}  & 14.3  & 35.4$^{\dagger}$  & 67.7$^{\dagger}$  & \underline{15.8}  & \underline{12.6}  & 34.2$^{\dagger}$  & 67.4$^{\dagger}$  \\
    \UR (CL 20\%) & \underline{24.6}$^{\dagger}$  & \underline{17.4}  & \underline{14.4} & \underline{35.6}$^{\dagger}$  & \underline{67.8}$^{\dagger}$  & \underline{15.8}  & \underline{12.6}  & \underline{34.3}$^{\dagger}$  & \underline{67.5}$^{\dagger}$  \\
 
    \midrule
    REPLUG (CL 100\%) & 25.0  & 17.2  & 14.2  & 35.8  & 67.8  & 15.8  & 12.6  & 34.4  & 67.5  \\
    \US (CL 100\%) & \textbf{\underline{25.6}}$^{+}$ & \textbf{\underline{17.8}} & \textbf{\underline{14.8}} & \textbf{\underline{36.0}} & \textbf{\underline{68.0}}$^{+ \dagger}$  & \textbf{\underline{16.2}} & \textbf{\underline{12.9}} & \textbf{\underline{34.6}}$^{+ \dagger}$  & \textbf{\underline{67.7}}$^{+ \dagger}$  \\
    \UR (CL 100\%) & 25.5$^{+}$  & 17.7  & 14.7  & 35.9  & \textbf{\underline{68.0}}$^{+ \dagger}$ & \textbf{\underline{16.2}}  & \textbf{\underline{12.9}}  & \textbf{\underline{34.6}}$^{+ \dagger}$ & \textbf{\underline{67.7}}$^{+ \dagger}$  \\
    \bottomrule
    \end{tabular}
\caption{
RAG performance (\%) of different retrievers (RetroMAE backbone) trained with various MS MARCO annotations on MS MARCO QA dataset. 
The symbols $^{+}$, $^{-}$, and $^{\dagger}$ are defined in Table \ref{tab:diff_annotation}. \textbf{Bold} and \underline{underline} are also defined in Table \ref{tab:diff_annotation}. The official BLEU evaluation for MS MARCO QA targets the entire queries, not individual queries, thus no significance tests are conducted.}
\label{tab:rag_in_domain}
\end{table*}

\begin{table*}[t]
\centering
\small
\setlength\tabcolsep{3pt}
\begin{tabular}{lllllllllll}
\toprule
    \multicolumn{1}{l}{\multirow{3}[6]{*}{Annotation}} & \multicolumn{5}{c}{NQ}        & \multicolumn{5}{c}{HotpotQA} \\
    \cmidrule(r){2-6}   \cmidrule(r){7-11} 
    & \multicolumn{1}{c}{\multirow{2}[4]{*}{Recall}} & \multicolumn{2}{c}{Llama} & \multicolumn{2}{c}{Qwen} & \multicolumn{1}{c}{\multirow{2}[4]{*}{Recall}} & \multicolumn{2}{c}{Llama} & \multicolumn{2}{c}{Qwen} \\
    \cmidrule(r){3-4}\cmidrule(r){5-6} \cmidrule(r){8-9}  \cmidrule(r){10-11}    
    &   & \multicolumn{1}{l}{EM} & \multicolumn{1}{l}{F1} & \multicolumn{1}{l}{EM} & \multicolumn{1}{l}{F1} &   & \multicolumn{1}{l}{EM} & \multicolumn{1}{l}{F1} & \multicolumn{1}{l}{EM} & \multicolumn{1}{l}{F1} \\
    
    \midrule
    Human & 56.7  & 42.8  & 56.4  & 43.6  & 57.9  & 54.8  & 31.5  & 42.6  & 38.6  & 50.7  \\
    \midrule
    REPLUG & 46.2$^{-}$  & 41.1$^{-}$  & 53.7$^{-}$  & 41.6$^{-}$  & 55.0$^{-}$  & 53.3$^{-}$  & 30.6$^{-}$  & 41.6$^{-}$  & 38.0 & 50.0$^{-}$ \\
    \US & 61.1$^{+ \dagger}$  & 44.4$^{+ \dagger}$  & 58.8$^{+ \dagger}$  & 44.9$^{\dagger}$  & 59.8$^{+ \dagger}$  & 55.8$^{+ \dagger}$  & \textbf{\underline{31.9}}$^{\dagger}$ & \underline{43.2}$^{\dagger}$  & \textbf{\underline{39.0}}$^{\dagger}$  & \underline{51.1}$^{\dagger}$  \\
    \UR & \textbf{\underline{62.0}}$^{+ \dagger}$  & \textbf{\underline{45.4}}$^{+ \dagger}$ & \textbf{\underline{59.8}}$^{+ \dagger}$ & \textbf{\underline{45.9}}$^{+ \dagger}$ & \textbf{\underline{60.0}}$^{+ \dagger}$ & \underline{55.9}$^{+ \dagger}$  & 31.4$^{\dagger}$  & 43.0$^{\dagger}$  & 38.7  & 51.0$^{\dagger}$ \\
    \midrule

    REPLUG (CL 20\%) &  55.0$^{-}$   & 43.3  & 56.9  & 44.7  & 58.4  &  56.5$^{+}$  & 31.3  & 42.6  &  38.6  & 50.7  \\
    \US (CL 20\%) & \underline{59.8}$^{+ \dagger}$  & 43.4  & 58.0$^{+}$  & 44.9$^{+}$  & 59.3$^{+}$  & \underline{56.2}$^{+}$  & \textbf{\underline{31.9}}  & \underline{43.0}  & 38.8 & 51.0 \\
    \UR (CL 20\%) & 59.7$^{+ \dagger}$  & \underline{44.7}$^{+}$  & \underline{58.9}$^{+ \dagger}$  & \underline{45.6}$^{+}$  & \underline{59.7}$^{+ \dagger}$  & \underline{56.2}$^{+}$  & 31.5  & 42.9  & \textbf{\underline{39.0}} & \textbf{\underline{51.3}}  \\
    \midrule
    
    REPLUG (CL 100\%) &   58.2$^{+}$    & 43.5  & 57.2  & 45.3$^{+}$  & 59.2$^{+}$  &  \textbf{\underline{57.1}}$^{+}$   & \underline{31.8}  & \textbf{\underline{43.3}}$^{+}$ & \underline{38.8}  & \underline{51.1} \\
    \US (CL 100\%) & \underline{59.9}$^{+ \dagger}$  & 43.7  & 57.5  & \underline{45.4}$^{+}$  & \underline{59.8}$^{+}$  & 56.6$^{+}$  & 31.7  & 43.2  & 38.7  & 50.8  \\
    \UR (CL 100\%) & 59.4$^{+ \dagger}$  & \underline{43.8}  & \underline{57.8}$^{+}$  & 45.0$^{+}$  & 59.1$^{+}$  & 56.0$^{+}$  & 31.4 & 42.9  & 38.4  & 50.7  \\
    \bottomrule
\end{tabular}
\caption{
RAG performance (\%) of different retrievers (RetroMAE backbone) trained with various MS MARCO annotations on the NQ and HotpotQA datasets. 
The symbols $^{+}$, $^{-}$, and $^{\dagger}$ are defined in Table \ref{tab:diff_annotation}. \textbf{Bold} and \underline{underline} are also defined in Table \ref{tab:diff_annotation}. ``Llama'' and ``Qwen'' are ``Llama-3.1-8B'' and ``Qwen-2.5-32B-Int8'', respectively.}
\label{tab:rag_out_of_domain}
\end{table*}

\section{Experimental Results}
\label{sec:exp_results}

\subsection{Retrieval Performance}
\label{subsec:retrieval_perf}

\heading{In-domain Results} 
Table~\ref{tab:diff_annotation} shows the overall in-domain retrieval performance. Main findings include: 
1) On human-labeled test sets, models trained with human relevance annotations perform better than using LLM annotations alone, and they are both better than training with downstream task performance (REPLUG). 
2) When combining 20\% human labels, the model performance of \US and \UR has no significant difference with using all the human annotations. This means that \US and \UR can save about 80\% human effort on this dataset to achieve similar performance. 
3) With 100\% human annotations, \US and \UR can achieve significant improvements over using human annotations alone, which confirms the efficacy of our annotation and training strategy as a data augmentation approach. 
4) Regarding both human and GPT-4 annotated golden labels, \US and \UR significantly outperform models trained with human annotations alone, indicating their potential in a fairer setting.

\heading{Out-of-domain (OOD) Results} 
Table~\ref{tab:beir_zero_shot} and Table~\ref{app:tab_contriever_beir} (in Appendix~\ref{app:contriever_beir}) report the zero-shot retrieval performance of RetroMAE and Contriever trained with different annotations.
We observe the following: 
1) Both \US and \UR exhibit superior out-of-domain (OOD) performance compared to retrievers trained solely on MS MARCO human annotations. This indicates that reliance on MS MARCO human labels may lead to model overfitting to the corpus. The fact that \US outperforms \UR and it utilizes more LLM annotations than \UR, as shown in Table \ref{tab:annotationRecall}, further supports this observation. 
2) When incorporating 20\% or 100\% human labels during training, the OOD retrieval performance decreases compared to not using them, reinforcing the first point.
These findings suggest a trade-off between in-domain and OOD retrieval performance, which can be adjusted by varying the mix of MS MARCO human labels with LLM annotations.

\subsection{RAG Performance}

\heading{In-domain Results}
In Table~\ref{tab:rag_in_domain}, we present the RAG performance on MS MARCO QA using passages from retrievers (based on RetroMAE) compared in Section~\ref{subsec:retrieval_perf} for RAG. The findings are consistent with the first three conclusions regarding in-domain retrieval discussed in \ref{subsec:retrieval_perf}, which is expected as more accurate retrieval enhances generation. 
This confirms that \US and \UR can significantly reduce human annotation efforts while maintaining comparable RAG performance. Notably, REPLUG performs the poorest among the methods, differing from results in \citet{shi2024replug}. This discrepancy could arise because we used REPLUG for static utility annotation, whereas the original paper iteratively updated retrievers based on generation performance for RAG.

\heading{OOD Results}
Similarly, we assess the RAG performance based on MS MARCO-trained retrievers on NQ and HotpotQA.
Results are shown in Table~\ref{tab:rag_out_of_domain}. Key findings include:
1) \US and \UR consistently yield the best RAG performance across most generators and datasets (particularly on NQ), highlighting the potential of utility-focused LLM annotation in initializing QA systems. 
2) On NQ, the best RAG performance is observed when no human annotations are used, mirroring the retrieval performance trend across many BEIR datasets (in Table \ref{tab:beir_zero_shot}). In contrast, on HotpotQA, retrieval performance is improved when human labels are used, while RAG is not enhanced. These results suggest that human annotations do not significantly benefit \US and \UR for OOD RAG.

\section{Further Analysis} 
\label{sec:further_analysis}

\heading{Comparison of Strategy Variants}
Table \ref{tab:ana_llm_label} compares the variants of our annotation method and training strategies regarding the retrieval performance on MS MARCO. The default setting for each component when using LLM annotations for training is Qwen, \US, and \orfunction. 
Key findings are: 1) Within the same GPU memory, the quantized version of larger LLMs has better capacity than smaller ones (Qwen better than LLama); 2) \US and \UR lead to better performance than \RS, indicating stricter annotation criterion is needed; 3) When multiple positives exist, \orfunction achieves the best performance, indicating its robustness to potential noise introduced by LLM annotations. 4) When integrating human annotations, training with higher-quality human annotations at last outperforms optimizing towards the union of positives from humans and LLMs.

\begin{table}[t]
\centering
\small
\setlength\tabcolsep{1.3pt}
\begin{tabular}{l l cc}
    \toprule
    Method/Component  & Variants  & MRR@10  & R@1000 \\
    
    \midrule
    Human & -  & 38.6  & 98.6  \\
    \midrule

    \multirow{2}[1]{*}{LLM Annotator} & Llama-8B & 33.0  & 97.4  \\
          & Qwen-32B-Int8 & \textbf{35.3}  & \textbf{97.7}  \\
    \midrule
    
    \multirow{3}[2]{*}{Annotation Strategy \text{}} 
          &  \RS  & 33.5  & \textbf{97.9}  \\
          & \US  & 35.3  & 97.7  \\
          & \UR  & \textbf{35.7}  & 97.8  \\
    \midrule
    
    \multirow{3}[2]{*}{Training Loss} & \randfunction & 34.5  & \textbf{97.9}  \\
          & \andfunction & 34.0  & 97.5  \\
          & \orfunction & \textbf{35.3}  & 97.7  \\
    \midrule
    
    
    \multirow{2}[1]{*}{+20\% Human Labels \text{}} & Positive Union  & 33.2  & 97.2 \\
     & CL & \textbf{38.2}  & \textbf{98.5} \\
    
    \bottomrule
\end{tabular}
\caption{
Controlled experiments using LLM annotations for training. 
See Appendix \ref{app:implementation} for detailed settings.
}
\label{tab:ana_llm_label}
\end{table}

\begin{figure}[htbp]
    \centering
    \includegraphics[width=\linewidth]{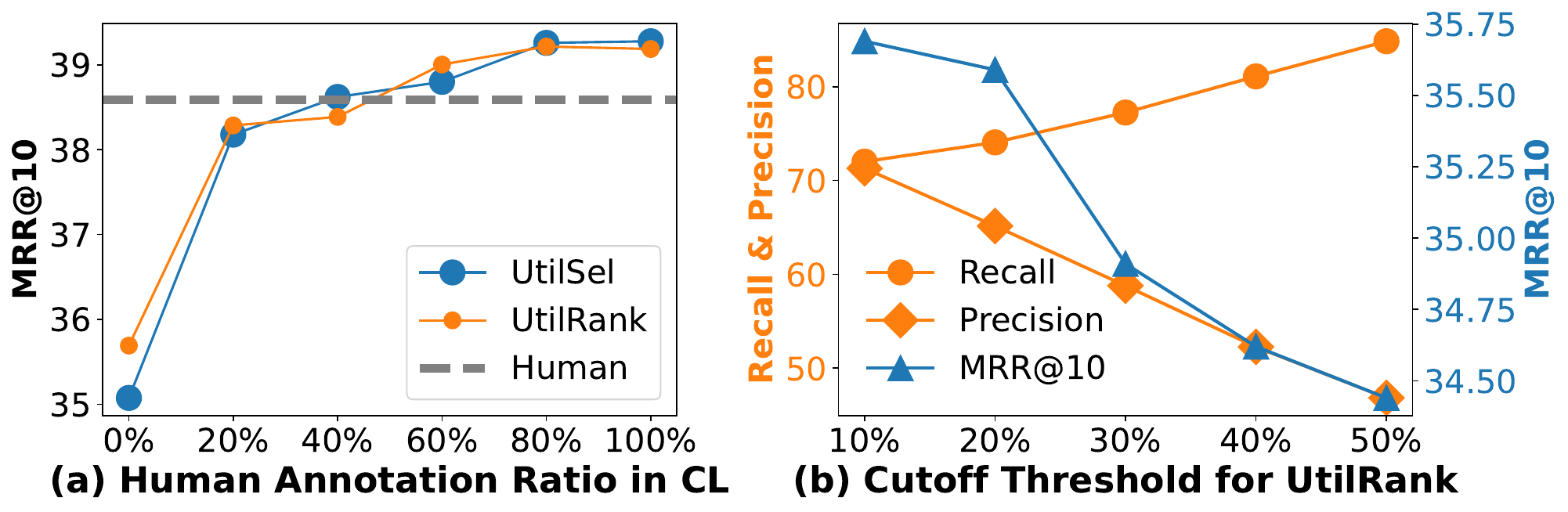}
    \caption{(a): Retrieval performance (\%) with different human annotation ratios in curriculum learning; (b): Annotation quality evaluation (\%) and retrieval performance (\%) with different thresholds for \UR.}
    \label{fig:cl_ratio}
\end{figure}

\heading{Human Annotation Ratio in CL} 
Figure \ref{fig:cl_ratio} shows the retrieval performance of using different ratios of human annotations in CL on the MS MARCO Dev set. It indicates that the in-domain retrieval performance increases with more human-labeled data used in CL.

\heading{Cutoff Threshold for \UR}
As illustrated in Figure~\ref{fig:cl_ratio}, smaller thresholds result in higher precision while lower recall regarding human-labeled ground truth, and better in-domain retrieval performance. This again confirms that stricter criteria and fewer positives lead to better in-domain retrieval performance. It is not surprising since this results in a positive-to-negative ratio more closely aligned with the distribution encountered during inference. 

\section{Conclusion} 
In this work, we explore the use of LLMs to annotate large-scale retrieval training datasets with a focus on utility to reduce dependence on costly human annotations. 
Experiments show that retrievers trained with utility annotations outperform retrievers trained with human annotations in out-of-domain settings on both retrieval and RAG tasks. 
Furthermore, we investigate combining LLM annotations with human annotations by curriculum learning. 
Interestingly, with only 20\% of human annotations, the performance of the retriever trained on utility annotations has no significant decline over full human annotations. 
Moreover, with 100\% human annotations yields a significant improvement over training solely on human annotations. 
This highlights the effectiveness of LLM-generated annotations as weak supervision in the early stages of training. 
Our study offers a comprehensive approach to utilizing LLM annotations for initializing QA systems on new corpora.


\section{Limitations}

There are several limitations that should be acknowledged:
1) Our annotation pool is constructed using human-annotated positives and hard negatives retrieved by other models. It may not fully reflect real-world annotation scenarios, where candidates are typically retrieved using unsupervised methods like BM25 or retrievers trained on other data. We analyze the impact of including human-labeled positives in Appendix~\ref{app:human_must_yes_no}.
2) Due to time and resource constraints, we did not adopt stronger LLMs for annotation, though they may offer further improvements. 
Moreover, our annotations are limited to MS MARCO, a standard dataset for retrieval. Extending this approach to RAG datasets like NQ remains a promising direction, as our analysis suggests that similar trends would likely hold. 
To further investigate this, we leverage a SOTA open-source LLM, Qwen3-32B \cite{yang2025qwen3}, for annotation on the NQ dataset. 
The results are shown in Appendix~\ref{app:additional_dataset}. 
The conclusion is that LLM annotations can achieve comparable performance to relevance annotations based on human answers.

\section{Ethics Statement}
Our research does not rely on personally identifiable information. 
All datasets, pre-trained IR models, and LLMs used in this study are publicly available, and we have properly cited all relevant sources. 
We firmly believe in the principles of open research and the scientific value of reproducibility. To this end, we have made all our code, data, and trained models associated with this paper publicly available on GitHub. 


\section*{Acknowledgements}
This work was supported by several grants, including the National Natural Science Foundation of China (Grant Nos. 62441229 and 62302486), the Innovation Funding of ICT CAS (Grant No. E361140 and No.E561010), the CAS Special Research Assistant Funding Project, the project under Grant No. JCKY2022130C039, and the Strategic Priority Research Program of the CAS (Grant No. XDB0680102). 
\bibliography{custom}
\appendix

\begin{table*}[t]
\centering
\small
\renewcommand{\arraystretch}{0.7}
\setlength\tabcolsep{2pt}
\begin{tabular}{lllllll}
    \toprule
    \multirow{2}[3]{*}{Annotation} & \multicolumn{4}{c}{Human Test} & \multicolumn{2}{c}{Hybrid Test} \\
    \cmidrule(r){2-5}  \cmidrule(r){6-7}         & \multicolumn{1}{c}{MRR@10} & \multicolumn{1}{c}{Recall@1000} & \multicolumn{1}{c}{DL19 (NDCG@10)} & \multicolumn{1}{c}{DL20 (NDCG@10)} & \multicolumn{1}{c}{MRR@10} & \multicolumn{1}{c}{NDCG@10} \\
    \midrule
    Human & 38.6  & 98.6  & 68.2  & 71.6  & 83.7  & 63.1  \\
    \midrule
    \textit{Exclusion} (0\%) & 31.2$^{-}$  & 97.1$^{-}$  & 64.6  & 70.2  & 84.5  & 63.3  \\
    \textit{Exclusion} (CL 20\%) & 37.4$^{-}$  & 98.5  & 70.5  & 69.4  & 84.2  & 63.0$^{-}$  \\
    \textit{Exclusion} (CL 30\%) & 38.2  & 98.5  & 69.3  & 70.4  & 85.0  & 64.2$^{+}$  \\
    \midrule
    \textit{Random} (0\%) & 35.3$^{-}$  & 97.7$^{-}$  & 68.0  & 71.0  & 87.5$^{+}$  & 65.8$^{+}$  \\
    \textit{Random} (CL 20\%) & 38.2  & 98.5  & 69.6  & 71.4  & 83.4  & 65.5$^{+}$  \\
    \midrule
    \textit{Inclusion} (0\%) & 36.1$^{-}$  & 98.1$^{-}$  & 69.0  & 71.3  & 87.7  & 66.7$^{+}$  \\
    \textit{Inclusion} (CL 20\%) & 38.2  & 98.6  & 70.9  & 70.7  & 84.2  & 64.6$^{+}$  \\
    \bottomrule
\end{tabular}
\caption{Retrieval performance (\%) with different \US annotation labels on whether human-annotated relevant passage is included or not during training (i.e., \textit{Exclusion}, \textit{Random}, \textit{Inclusion}) using RetroMAE backbone.  ``$^{+}$'' and ``$^{-}$'' indicate significant improvements and decrements over Human using a two-sided paired t-test ($p<0.05$).
}
\label{tab:app:must_yes_no}
\end{table*}

\begin{table*}[htbp]
\centering
\small
\renewcommand{\arraystretch}{0.7}
\setlength\tabcolsep{2pt}
\begin{tabular}{lcccccccc}
    \toprule
    \multirow{2}[4]{*}{Dataset} & \multicolumn{1}{c}{\multirow{2}[4]{*}{Human}} & \multicolumn{2}{c}{\textit{Random}} & \multicolumn{3}{c}{\textit{Exclusion}} & \multicolumn{2}{c}{\textit{Inclusion}} \\
    \cmidrule(r){3-4}  \cmidrule(r){5-7} \cmidrule(r){8-9}            &       & 0\%   & (CL, 20\%) & 0\%   & (CL, 20\%) & (CL, 30\%) & 0\%   & (CL, 20\%) \\
    \midrule
    DBPedia & 36.0  & 38.0  & 37.4  & \textbf{39.0}  & 37.3  & 37.1  & \underline{38.8}  & 37.0  \\
    FiQA  & 29.7  & \underline{32.6}  & 32.1  & 30.1  & \textbf{32.8}  & 31.2  & \underline{32.6}  & 32.3  \\
    NQ    & 49.2  & \underline{53.5}  & 51.4  & 52.2  & 51.0  & 51.8  & \textbf{53.7}  & 51.0  \\
    HotpotQA & 58.4  & 59.6  & 60.0  & 59.1  & \textbf{60.5}  & \underline{60.4}  & 59.9  & 60.3  \\
    NFCorpus & 32.8  & 33.9  & 34.2  & \textbf{34.4}  & \underline{34.3}  & 33.4  & 34.1  & \textbf{34.4}  \\
    T-COVID & 63.4  & 66.1  & 65.0  & 60.3  & \underline{67.4}  & 66.1  & 65.1  & \textbf{67.6}  \\
    Touche & 24.2  & \textbf{28.5}  & 24.7  & 25.3  & \underline{26.5}  & 26.2  & 25.0  & 26.2  \\
    CQA   & 32.2  & 32.3  & \underline{33.9}  & 32.2  & \textbf{34.7}  & 33.4  & 32.4  & 33.8  \\
    ArguAna & 30.5  & 34.1  & 36.4  & \textbf{39.3}  & \underline{38.5}  & 36.4  & 37.9  & 36.8  \\
    C-FEVER & 18.0  & \textbf{19.5}  & 16.5  & \underline{19.3}  & 17.2  & 16.7  & 18.3  & 17.2  \\
    FEVER & 66.6  & 73.8  & 69.9  & 69.9  & \underline{71.4}  & \textbf{71.6}  & 71.0  & 71.2  \\
    Quora & \underline{86.2}  & 85.4  & 86.1  & 84.9  & \underline{86.2}  & \textbf{86.3}  & 85.8  & \underline{86.2}  \\
    SCIDOCS & 13.4  & 14.3  & \underline{14.4}  & \textbf{14.5}  & 14.2  & 14.1  & 14.3  & 14.1  \\
    SciFact & 63.1  & 62.8  & \textbf{64.2}  & 62.9  & \underline{63.9}  & \textbf{64.2}  & 63.2  & 63.2  \\
    \midrule
    Avg & 43.1  & \underline{45.3}  & 44.7  & 44.5  & \textbf{45.4}   & 44.9  & 45.2  & 45.1  \\
    \bottomrule
\end{tabular}
\caption{Zero-shot retrieval performance (NDCG@10, \%) with different \US annotation labels on whether human-annotated relevant passage is included or not during training using RetroMAE backbone.}
\label{app:tab:beir_human_must_yes_no}
\end{table*}

\section{Preliminary}
\label{app:preliminaries}

\subsection{Typical Dense Retrieval Models}
Dense retrieval models primarily employ a two-tower architecture of pre-trained language models, i.e.,$\mathcal{R}_{q}(\cdot)$ and $\mathcal{R}_{d}(\cdot)$, to encode query and passage into fixed-length dense vectors. 
The relevance between the query $q$ and passage $d$ is $s(q, d)$, i.e., 
\begin{equation}
s(q, d)=f<\mathcal{R}_{q}(q), \mathcal{R}_{d}(d)>, 
\end{equation}
where $f<\cdot>$ is usually implemented as a simple metric, e.g., dot product and cosine similarity. $\mathcal{R}_{q}(\cdot)$ and $\mathcal{R}_{d}(\cdot)$ usually share the parameters. 

\subsection{Downstream Task Performance as Utility Score} 
\label{app:ds-utility} 
Considering the downstream task for the retriever, i.e., RAG, the goals of the retriever and generator in RAG are different and can be mismatched. 
To alleviate this issue, the utility of retrieval information $f_u(q,d,a)$, where $a$ is the ground truth answer, enables the retriever to be more effectively alignment with the generator. 
$f_u(q,d,a)$ mainly has two ways: 
directly model how likely the candidate passages can generate the ground truth answer \cite{shi2024replug}, i.e., $P(a|q,d)$, 
which computes the likelihood of the ground truth answer; 
and measure the divergence of model output $LLM(q,d)$ and the answer $a$ using evaluation metrics \cite{zamani2024stochastic}, e.g., EM, i.e., $EM(a, LLM(q,d))$. 
Given the query $q$ and candidate passage list $D=[d_1, d_2, ..., d_n]$, where $n=|D|$. 
The optimization of the retriever is to minimize the KL divergence between the relevance distribution $R=\{s'(q,d_i)\}_{i=1}^N$, where $s'(q,d_i)$ is the relevance $s(q,d_i)$ from retriever after softmax operation, and utility distribution $U=\{f'_u(q,d_i, a)\}_{i=1}^N$, where $f'_u(\cdot)$ is the utility function $f_u(\cdot)$ from generator after softmax: 
\begin{equation}
\small
KL(U||R)=\sum_{i=1}^{N}U(d_i)log(\frac{U(d_i)}{R(d_i)}).
\end{equation}


\begin{table*}
    \centering
    \begin{tabular}{lccccc}
    \toprule
    Annotation &	Top20&	Top40	&Top60&	Top80&	Top100  \\
    \midrule
        Human (First1LH)	& 81.9& 	85.0& 	86.5	& 87.0 & 	87.8 \\
UtilSel (First1LH)& 	81.2& 	84.5	& 86.4	& 87.3	& 88.2 \\
UtilSel (SumMargLH)	& 81.6 & 84.8 & 86.4 & 	87.2	& 88.0 \\
\bottomrule
    \end{tabular}
    \caption{Retrieval performance (\%) of different annotation methods on the NQ dataset using Qwen3-32B annotation. All three groups of results do not have significant differences with p < 0.05. }
    \label{tab:nq-data}
\end{table*}

\section{Additional Analyses of Training Strategies} 
\label{app:additional_analyses}

\subsection{Impact of Human Annotated Positive}
\label{app:human_must_yes_no}
When generating LLM annotations, the model relies on a pool that includes human-annotated positives and retrieved negatives. 
To examine whether the presence of human-annotated positives in this pool influences retriever training, we compare three strategies:
\begin{enumerate*}
    \item \textit{Random}: The default strategy in our main experiments. 
    Positives and negatives of each query are randomly sampled from all LLM annotationed positive and negative instances, respectively, without distinguishing human-annotated examples during retriever training. 
    \item \textit{Exclusion}: Human-annotated positives are explicitly excluded during retriever training. Sepcifically, passages for each query during training are randomly selected from the LLM annotations which excluding human-annotated passages. 
    \item \textit{Inclusion}: Human-annotated positives for each query are always included during training, the rest are randomly sampled from the remaining LLM-labeled passages.
\end{enumerate*}

Tables~\ref{tab:app:must_yes_no} and \ref{app:tab:beir_human_must_yes_no} report in-domain and out-of-domain retrieval performance under three sampling strategies.
We draw three main observations:
\begin{enumerate*}[leftmargin=*,itemsep=0pt,topsep=0pt,parsep=0pt]
    \item Excluding human positives substantially degrades performance, highlighting their importance as high-quality signals. 
    As shown in Table~\ref{tab:annotationRecall}, LLMs consistently recall human positives, indicating their strong alignment with human judgments. 
    Removing them reduces annotation quality and hinders retriever training. 
    Conversely, explicitly including human positives in each batch yields the best results.
    
    \item Despite the initial performance gap under the \textit{Exclusion} setting, introducing 30\% human-labeled data in the second stage of curriculum learning effectively closes the gap. 
    The resulting model performs on par with those trained using the full human set, suggesting that LLM-generated negatives and non-human positives still provide valuable learning signals when combined with even partial human supervision. 
    
    \item For OOD performance, the \textit{Exclusion} setting outperforms the model trained purely on human labels, consistent with the main findings under the \textit{Random} setting. 
\end{enumerate*}

\subsection{Positive Sampling Strategies}
\label{app:positive_sampling}
LLM annotations might yield multiple positive instances. 
If the loss function is \orfunction or \andfunction, for their positive selection during training for each query, we devised three strategies: 
\begin{enumerate*}[leftmargin=*,itemsep=0pt,topsep=0pt,parsep=0pt]
    \item \textit{Pos-one}: randomly select one annotated positive instance, and sample the remaining examples from other positives and negatives; 
    
    \item  \textit{Pos-avg}: compute the average number of positive instances per query from LLM annotations, then sample this number of positives randomly for each query, with the rest sampled from negatives;
    
    \item  \textit{Pos-all}: include all annotated positive instances whenever available, and sample the remaining examples from negatives (ensuring at least one negative instance is included).
\end{enumerate*} 

As shown in Table~\ref{app:tab_pos_sample}, these positive sampling strategies have limited effect on standard retriever training using LLM annotations, but show a more noticeable impact in the curriculum learning setting. 
This may be because human-labeled data typically contain fewer positive examples, making the \textit{Pos-one} strategy more aligned with their distribution than \textit{Pos-all}, thereby reducing distribution mismatch during curriculum learning.

\begin{table}[htbp]
\centering
\small
\renewcommand{\arraystretch}{0.8}
\setlength\tabcolsep{2pt}
\begin{tabular}{lcc}
    \toprule
    Sampling  & MRR@10  & Recall@1000 \\
    \midrule
    
    \textit{Pos-one} & 35.1  & 97.7 \\
    \textit{Pos-avg} & 35.1   & 97.7 \\
    \textit{Pos-all} & 35.3  & 97.7  \\
    \midrule
        
    \textit{Pos-one} (CL) & 38.2  & 98.5 \\
    \textit{Pos-all} (CL) & 37.8 & 98.5 \\
    \bottomrule
\end{tabular}
\caption{
Effect of positive sampling strategies in training, evaluated under the \US annotations.
}
\label{app:tab_pos_sample}
\end{table}

\section{Additional Analyses on NQ Dataset}
\label{app:additional_dataset}
We conduct annotations on a more realistic scenario for NQ to show the efficacy of our utility-focused annotation pipeline:
(a) We constructed annotation candidates using unsupervised (BM25) and two out-of-domain retrievers trained on MS MARCO, i.e., our UtilSel trained on MS MARCO (RetroMAE backbone)  and LLM-QL \cite{zhang2025unleashing}.
(b) We annotated candidates via Qwen3-32B \cite{yang2025qwen3} (a state-of-the-art open-source LLM) to build the training set.
We trained retrievers using RetroMAE as the backbone with different annotations on NQ, including the original relevance annotations based on human answers, and our LLM annotations, as shown in Table \ref{tab:nq-data}. 
Following the standard practice for NQ \cite{karpukhin2020dense}, we used the First1LH setting (maximizing the likelihood of the first positive) for the original data, where only the first provided positive passage is used. 
For our LLM-annotated data, we experimented with both First1LH and SumMargLH loss. 
Our results demonstrate that our utility-focused LLM annotation approach can achieve similar performance compared to the original relevance annotation based on human-annotated answers, saving considerable manual labeling effort.




\begin{table*}[t]
    \centering
    \small
    \renewcommand{\arraystretch}{0.8}
   \setlength\tabcolsep{3.2pt}
    \begin{tabular}{lrrr rrr}
    \toprule
      &  \multicolumn{3}{c}{Retrieval}  & \multicolumn{3}{c}{RAG}  \\
      \cmidrule(r){2-4}   \cmidrule(r){5-7}  
     Datasets & MS MARCO Dev & TREC DL-19 & TREC DL-20 & MS MARCO-QA & NQ  & HotpotQA   \\ 
        \midrule
        \#Queries & 6980 & 43 & 54  & 6980 & 2255 & 7405 \\ 
        \#Rel.Passage per query & 1.1 & 95.4 & 66.8 & 1.1 & 1.2  & 2  \\ 
        \#Graded.Retrieval labels &  2 &  4 &  4 & 2 & 2 & 2 \\ 
    \bottomrule
    \end{tabular}
    \caption{Statistics of retrieval and RAG datasets. }
     \label{app:tab:datasets}%
\end{table*}

\section{Detailed Experimental Settings}

\subsection{Retrieval and RAG Datasets}
\label{app:dataset} 

\heading{Retrieval Datasets}
Three human-annotated test collections are used for in-domain retrieval evaluation: the MS MARCO Dev set \cite{nguyen2016ms}, which comprises 6980 queries, and TREC DL19/DL20 \cite{craswell2020overview, craswell2021overview}, which include 43 and 54 queries from MS MARCO Dev set. DL19 and DL20 have more human-annotated relevant passages, with each query having an average of around 95 and 67 positives, respectively. 
We further evaluate the zero-shot performance of our retrievers on 14 publicly available datasets from the BEIR benchmark, excluding MS MARCO \cite{nguyen2016ms}, which is used for training. The evaluation datasets include TREC-COVID \cite{voorhees2021trec}, NFCorpus \cite{boteva2016full}, NQ \cite{kwiatkowski2019natural}, HotpotQA \cite{yang2018hotpotqa}, FiQA \cite{maia201818}, ArguAna \cite{wachsmuth2018retrieval}, Touche \cite{bondarenko2020overview}, Quora, DBPedia \cite{hasibi2017dbpedia}, SCIDOCS \cite{cohan2020specter}, FEVER \cite{thorne2018fever}, Climate-FEVER \cite{diggelmann2020climate}, SciFact \cite{wadden2020fact}, and CQA \cite{hoogeveen2015cqadupstack}. 

\heading{RAG Datasets} 
For the in-domain setting, we use the MS MARCO QA dataset, which contains ground-truth answers for MS MARCO Dev queries on in-domain RAG evaluation. 
For the out-of-domain setting, we use two factoid question datasets in the BEIR benchmark for RAG evaluation: NQ \cite{kwiatkowski2019natural}, which consists of real questions issued to the Google search engine, and HotpotQA \cite{yang2018hotpotqa}, which consists of QA pairs requiring multi-hop reasoning gathered via Amazon Mechanical Turk. 
We used the queries with ground truth answers from 3,452 queries on NQ and then collected 2,255 queries for RAG evaluation.  
Table \ref{app:tab:datasets} shows detailed statistics of the in-domain retrieval datasets and all RAG datasets used in our work.

\subsection{Implementation Details} 
\label{app:implementation}
The retriever is trained for 2 epochs using the AdamW optimizer with a batch size of 16 (per device) and a learning rate of 3e-5.
Training is conducted on a machine with 8 $\times$ Nvidia A800 (80GB) GPUs. To ensure reproducibility of the single run, the random seed that will be set at the beginning of training using the default value. 
In the second stage of curriculum learning, the retriever is further trained for 1 epoch with the same hyper-parameters, except that the learning rate is re-initialized to 3e-5.

Unless otherwise specified, we use Qwen-2.5-32B-Int8 as the annotator, adopt the \orfunction loss with \US annotations, and apply the \textit{Pos-all} strategy for selecting positives. During curriculum learning, the positive sampling strategy is switched to \textit{Pos-one} (see Appendix~\ref{app:positive_sampling} for details). 
Due to the top 10\% ranked list of \UR containing an average of one positive, and \orfunction have no advantage in \UR, we use \randfunction loss for training under \UR.  

For RAG evaluation, the retrieved passages are directly fed to LLMs. We use top-1 passage for MS MARCO QA and top-5 passages for NQ and HotpotQA. The rationale for these choices is discussed in Appendix~\ref{app:top-k-in-rag}.

The original REPLUG \cite{shi2024replug} uses Contriever \cite{izacard2021contriever} and optimizes the retriever by aligning its relevance scores with LLM-derived utility scores via KL divergence. Our setup follows the overall REPLUG framework but differs in two key aspects: we adopt the same retriever backbone as in other experiments for fair comparison, and use static negatives during training instead of dynamically generated ones.

\begin{table*}[t]
\centering
\small
\renewcommand{\arraystretch}{0.8}
\setlength\tabcolsep{1.3pt}
\begin{tabular}{ccccccc ccccc}
    \toprule
    \multicolumn{1}{c}{\multirow{2}[4]{*}{Datasets}} & \multicolumn{1}{c}{\multirow{2}[4]{*}{Human}} & \multicolumn{1}{c}{\multirow{2}[4]{*}{REPLUG}}  & \multicolumn{1}{c}{\multirow{2}[4]{*}{\US}} & \multicolumn{1}{c}{\multirow{2}[4]{*}{\UR}} & \multicolumn{3}{c}{Curriculum Learning, 20\%} & \multicolumn{3}{c}{Curriculum Learning, 100\%} \\
    \cmidrule(r){6-8}   \cmidrule(r){9-11}
    &       &      &       & \multicolumn{1}{c}{} & \multicolumn{1}{c}{REPLUG} & \multicolumn{1}{c}{\US} & \multicolumn{1}{c}{\UR} & \multicolumn{1}{c}{REPLUG} & \multicolumn{1}{c}{\US} & \multicolumn{1}{c}{\UR} \\
    
    \midrule
    DBPedia & 34.5 & 26.6    & \textbf{37.3}  & \underline{36.9}  &  33.7  &  36.3  & 36.8  & 35.9  & 36.7  & 36.8  \\
    FiQA  & 28.3  & 22.5    & \textbf{30.1}  & 29.3  & 28.3  & 29.4  & \underline{29.6}  & 29.2  & 29.5  &  29.2  \\
    NQ    & 47.2  & 37.0    & \textbf{50.7} & \underline{50.7}  & 43.5  & 48.2  & 49.2  & 47.0  &  48.9  & 49.9  \\
    HotpotQA & 55.1  & 49.9    & 56.8  & 55.5  & 55.9  & 56.9  & 56.7  & \underline{56.9}  & \textbf{57.0}  & 56.9  \\
    NFCorpus & 30.4  & 28.0    & 31.3 & 31.1  & \underline{31.6}  & 31.3  & 30.9  & 31.5 & \textbf{31.8}  & 31.5  \\
    T-COVID & 49.9  & 26.9    & 53.4  & 55.1  & 34.8  & \underline{59.1}  & \textbf{62.2}  & 48.7  & 56.6  & 56.7 \\
    Touche & 20.1  & 14.7    & 23.7  & \textbf{26.6}  & 14.1  & 21.0  & \underline{26.0}  & 17.0  & 21.4  & 24.4  \\
    CQA  & 28.6  & 24.6    & 28.9  & 26.5  & \underline{29.9}  & \textbf{30.9} & 29.9  & 28.1  &  29.5  &  29.5  \\
    ArguAna  & 16.9  & 4.6    & 30.3  & 25.3  & 24.5  &  \textbf{34.2}  & \underline{32.3}  & 20.4  & 28.3  & 27.9  \\
    C-FEVER  & 14.3  & 8.9    & \textbf{20.0}  & 17.3 &  16.4 & 17.3  & 16.4  & \underline{17.5} & 17.4  & 17.2  \\
    FEVER  & 64.4  & 57.8    & 67.0  & \textbf{68.2} &  61.4  & 62.4  & 66.1  & 67.0  & 64.6  & \underline{67.6}  \\
    Quora  & 85.1  & 67.7    & 84.3  & 84.6  & 82.6  & 85.0  & 85.0  & 84.5  & \underline{85.5}  & \textbf{85.5} \\
    SCIDOCS  & 12.2  & 10.2    & 13.2  & 12.2  & \underline{13.2}  & \textbf{13.2} & 12.9  & 12.4  & 13.1  & 13.0  \\
    SciFact  & 61.7  & 54.8    & 64.8  & 61.6  & 62.2  & \underline{65.5}  & 62.9  & 63.7  & \textbf{65.7}  & 62.7  \\
    \midrule
    Average  & 39.2  & 31.0   & \underline{42.3}  & 41.5 & 38.0  & 42.2  & \textbf{42.6}  & 40.0  & 41.8  & 42.1  \\
    \bottomrule
\end{tabular}
\caption{Zero-shot retrieval performance (NDCG@10, \%) of different retrievers (Contriever backbone).}
\label{app:tab_contriever_beir}
\end{table*}

\begin{table*}[t]
\centering
\small
\renewcommand{\arraystretch}{0.8}
\setlength\tabcolsep{1.5pt}
\begin{tabular}{llccccccccc}
    \toprule
    \multirow{2}[3]{*}{Top-$k$} & \multirow{2}[3]{*}{Annotation} & \multirow{2}[3]{*}{Recall} & \multicolumn{4}{c}{Generator: LlaMa-3.1-8B} & \multicolumn{4}{c}{Generator: Qwen2.5-32B-Int8} \\
    \cmidrule(r){4-7}   \cmidrule(r){8-11}         &       &       & BLUE-3 & BLUE-4 & ROUGE-L & BERT-score & BLUE-3 & BLUE-4 & ROUGE-L & BERT-score \\
    
    \midrule
    \multirow{4}[2]{*}{Top 1} & Human & 24.7  & 17.2  & 14.2  & 35.7  & 67.8  & 15.8  & 12.6  & 34.3  & 67.4  \\
    & REPLUG & 21.7  & 15.7  & 12.9  & 33.8  & 66.7  & 14.7  & 11.6  & 32.4  & 66.2  \\
    & \US & 22.3  & 16.3  & 13.4  & 34.7  & 67.4  & 14.9  & 11.7  & 33.5  & 67.1  \\
    & \UR & 22.6  & 16.6  & 13.6  & 35.1  & 67.5  & 15.2  & 12.0  & 33.9  & 67.3  \\

    \midrule
    \multirow{4}[2]{*}{Top 5} & Human & 55.4  & 13.4  & 11.4  & 33.9 & 66.0  & 14.2 & 11.1 & 33.4 & 67.0 \\
    & REPLUG & 48.4  & 13.8  & 11.4  & 32.9  & 65.8  & 13.9  & 10.8  & 32.8  & 66.7  \\
    & \US & 51.5  & 14.3  & 11.8  & 33.3  & 66.1  & 13.7  & 10.7  & 33.0  & 66.8  \\
    & \UR & 51.6  & 14.4  & 11.9  & 33.3  & 66.1  & 13.8  & 10.7  & 32.9  & 66.8  \\
    
    \bottomrule
\end{tabular}
\caption{RAG performance with different top-$k$ on MS MARCO QA dataset (RetroMAE backbone).}
\label{app:tab:top-k}
\end{table*}

\subsection{Evaluation Metrics} 
\label{app:evaluation}
To evaluate retrieval performance, we employ three standard metrics: Mean Reciprocal Rank (MRR) \cite{craswell2009mean}, Recall and Normalized Discounted Cumulative Gain (NDCG) \cite{jarvelin2002cumulated}.  
To evaluate RAG performance, we adopt two different approaches based on the nature of the datasets:
\begin{enumerate*}[leftmargin=*,itemsep=0pt,topsep=0pt,parsep=0pt]
    \item For datasets that include non-factoid QA, such as MS MARCO, we evaluate answer generation performance using ROUGE \cite{lin2004rouge}, BLEU \cite{papineni2002bleu} \footnote{\url{https://github.com/microsoft/MSMARCO-Question-Answering/tree/master/Evaluation}}, and BERT-Score \cite{zhang2019bertscore} \footnote{We use the best model for BERT-Score: (\url{https://huggingface.co/microsoft/deberta-xlarge-mnli})}.
    
    \item For factoid QA datasets, such as NQ and HotpotQA, we use Exact Match (EM) and F1 score as main metrics. 
\end{enumerate*}


\begin{table*}[t]
\small
\centering
\renewcommand{\arraystretch}{0.7}
\setlength\tabcolsep{1pt}
\begin{tabular}{lcccc ccc}
    \toprule
    \multicolumn{1}{c}{\multirow{2}[2]{*}{Method}} & \multicolumn{1}{c}{\multirow{2}[2]{*}{Pre-training}} & \multicolumn{1}{c}{\multirow{2}[2]{*}{Hard Negatives}}   & \multicolumn{2}{c}{Dev} & \multicolumn{1}{c}{DL19} & \multicolumn{1}{c}{DL20} \\
    \cmidrule(r){4-5}   \cmidrule(r){6-6}  \cmidrule(r){7-7}
     &  &   & \multicolumn{1}{c}{M@10} & \multicolumn{1}{c}{R@1000} & \multicolumn{1}{c}{N@10} & \multicolumn{1}{c}{N@10} \\
    
    \midrule
    BM25 \cite{lin2021pyserini} & No & -  & 18.4  & 85.3  & 50.6  & 48.0  \\ 
    \midrule
    
    DPR \cite{karpukhin2020dense} & No & Static(BM25)  & 31.4 & 95.3 & 59.0 & - \\
    Condenser \cite{gao2021condenser} & Yes & Static(BM25)  & 33.8  & 96.1  & 64.8 & - \\ 
    RetroMAE \cite{xiao2022retromae} & Yes & Static(BM25)  & 35.5  & 97.6  & - & - \\  
     \midrule
     
    ANCE \cite{xiong2020approximate} & No & Dynamic  & 33.0  & 95.9  & 64.8  & - \\ 
    ADORE \cite{zhan2021optimizing} & No & Dynamic   & 34.7  & - & 68.3  & - \\ 
    CoCondenser \cite{gao2021unsupervised} & Yes & Dynamic   & 38.2  & 98.4  & 71.2  & 68.4  \\  
    SimLM \cite{wang2022simlm} & Yes & Dynamic  & 39.1  & 98.6  & 69.8  & 69.2  \\ 

    \midrule
    RetroMAE & Yes & Static(CoCondenser+BM25)  & 38.6 & 98.6 & 68.2 & 71.6 \\
    Contriever & Yes & Static(CoCondenser+BM25)  & 35.6 & 97.6 & 68.5 & 67.9  \\  
    \bottomrule
\end{tabular}
\caption{Retrieval performance on MS MARCO (measured by MRR@10, Recall@1000, NDCG@10).}
\label{tab:passage_retrieval}
\end{table*}

\begin{table*}[t]
\centering
\small
\renewcommand{\arraystretch}{0.7}
\setlength\tabcolsep{1.2pt}
\begin{tabular}{ccccc}
    \toprule
    \multicolumn{1}{c}{\multirow{2}[2]{*}{Datasets}} & \multicolumn{1}{c}{Static(BM25)}  &  \multicolumn{1}{c}{Dynamic}  & \multicolumn{2}{c}{Static(CoCondenser+BM25)}  \\
    \cmidrule(r){2-2} \cmidrule(r){3-3} \cmidrule(r){4-5}           
    &  RetroMAE \cite{xiao2022retromae} &   Contriever \cite{izacard2021contriever} &    RetroMAE   &    Contriever   \\

    \midrule
    MS MARCO &  -  &  40.7  &  45.2  &  42.1 \\
    \midrule
    DBPedia &  39.0  &  41.3 & 36.0  &  34.5 \\
    FiQA   & 31.6  & 32.9  & 29.7  &  28.3 \\
    NQ     &  51.8 &  49.8 & 49.2  &  47.2 \\
    HotpotQA & 63.5  & 63.8  & 58.4  &  55.1 \\
    NFCorpus & 30.8  &  32.8  & 32.8  &  30.4  \\
    T-COVID  &  77.2 &  59.6  & 63.4  &  49.9  \\
    Touche  & 23.7  &  23.0  & 24.2  &  20.1 \\
    CQA     &  31.7  &  34.5  & 32.2  &  28.6  \\
    ArguAna  &  43.3  &  44.6 & 30.5  &  16.9  \\
    C-FEVER  &  23.2  &  23.7  & 18.0  &  14.3  \\
    FEVER   &  77.4  &  75.8  & 66.6  &  64.4  \\
    Quora   &  84.7  &  86.5  & 86.2  &  85.1 \\
    SCIDOCS  &  15.0 &  16.5  & 13.4  &  12.2   \\
    SciFact  &  65.3 &  67.7  & 63.1  &  61.7  \\
    \midrule
    Average  & 47.0$^*$  &  46.6  &  43.1  &  39.2  \\
    \bottomrule
\end{tabular}
\caption{Zero-shot retrieval performance (NDCG@10, \%) on 14 BEIR datasets. MS MARCO is reported for reference but excluded from the average. Note that the original RetroMAE reports average performance over 18 datasets, while our reproduction only considers 14 publicly available datasets.}
\label{tab:appendix_zeroshot}
\end{table*}

\section{Supplementary Experimental Results}

\subsection{Zero-shot Retrieval Performance Using Contriever Backbone} 
\label{app:contriever_beir}

Table~\ref{app:tab_contriever_beir} compares the zero-shot retrieval performance of various retrievers built on the Contriever backbone.
All models are trained on MS MARCO using different annotation strategies, including human labels, REPLUG, utility-based annotations (\US and \UR), and corresponding curriculum learning variants.

\subsection{Top-$k$ in RAG} 
\label{app:top-k-in-rag}
Our top-$k$ choices in RAG evaluation reflect the characteristics of each dataset:
\begin{enumerate*}[leftmargin=*,itemsep=0pt,topsep=0pt,parsep=0pt]
    \item MS MARCO QA focuses primarily on non-factoid questions. As shown in Table \ref{app:tab:top-k}, including more passages tends to introduce irrelevant or verbose content, which lead to lower RAG performance. 
    Therefore, we use top-1 passage for evaluation.
    \item HotpotQA is a multi-hop factoid QA dataset, which naturally benefits from access to multiple supporting passages. Hence, we adopt top-5 passages (NQ also uses top-5 passages for consistency).
\end{enumerate*}
    
    

\subsection{Comparison with Reported Retrieval Results in Prior Work}
\label{app:retrieval_comparion}
In this section, we summarize the retrieval performance of several representative dense retrievers on MS MARCO and BEIR, based on results reported in their original papers.

Table~\ref{tab:passage_retrieval} shows performance on MS MARCO. Compared to the original results, our reproduction of RetroMAE shows slight differences. This can be attributed to the use of different hard negatives: while the original model used BM25-mined negatives, we employ a combination of BM25 and coCondenser negatives, which are more diverse and challenging. This leads to improved performance on MS MARCO by enhancing the ability to distinguish fine-grained semantic differences.

Table~\ref{tab:appendix_zeroshot} reports zero-shot performance on BEIR, measured by NDCG@10 across 14 datasets. Both RetroMAE and Contriever show a performance drop compared to their original results. We attribute this to the following factors:
\begin{enumerate*}[leftmargin=*,itemsep=0pt,topsep=0pt,parsep=0pt]
\item \textbf{For RetroMAE:} Our reimplementation uses stronger hard negatives during MS MARCO fine-tuning, which improves in-domain performance but may hinder generalization. Additionally, our model version is pre-trained on MS MARCO, whereas the original version was pre-trained on English Wikipedia and BookCorpus, which offer broader domain diversity and improved transferability.
\item \textbf{For Contriever:} The original paper uses only one hard negative per query and relies mainly on in-batch negatives, a strategy that mitigates overfitting and preserves generalization. In contrast, our setting introduces more difficult negatives, improving MS MARCO performance but leading to a drop on BEIR. Moreover, we adopt a unified setup for all models and use \texttt{[CLS]} pooling, whereas the original Contriever uses mean pooling, which may also contribute to the performance difference.
\end{enumerate*}

\vspace{-1mm}
\subsection{Further Analysis for \orfunction} 
\label{app:further_orfunction} 

\begin{table}[htbp]
\centering
\small
\renewcommand{\arraystretch}{0.7}
\setlength\tabcolsep{1.5pt}
\begin{tabular}{lcccc}
    \toprule
    \multirow{2}[4]{*}{Annotation} & \multirow{2}[4]{*}{Threshold} & \multirow{2}[4]{*}{Avg} & \multicolumn{2}{c}{Loss Function} \\
    \cmidrule{4-5}  
    &       &       & \orfunction & \randfunction \\
    \midrule
    \multirow{5}[2]{*}{\UR} & 10\%  & 1.0  & 35.6 & 35.7  \\
          & 20\%  & 1.3  & 35.4  & 35.6  \\
          & 30\%  & 1.7  & 35.1  & 34.9  \\
          & 40\%  & 2.3  & 34.7  & 34.6  \\
          & 50\%  & 3.0  & 34.6  & 34.4  \\
    \midrule
    \US & -  & 2.9  & 35.3  & 34.5  \\
    \bottomrule
\end{tabular}
\caption{Retrieval performance (MRR@10) on MS MARCO Dev using different loss functions across various annotation settings under RetroMAE backbone. ``Avg'' means the average number of positive instances.}
\label{app:tab:orfunction}
\end{table}

From Table~\ref{app:tab:orfunction}, we can observe the following:
1) When the number of positive instances is small, the advantage of \orfunction over \randfunction is limited. However, as the number increases, \orfunction generally yields better performance.
2) When the average number of positives is similar, \US outperforms \UR, suggesting that LLM-selected positives may be more effective than those chosen by thresholding.



\section{Efficiency and Cost} 
\label{app:efficiency}

According to \citet{gilardi2023chatgpt}, the cost of human annotation is approximately \$0.09 per annotation on MTurk, a crowd-sourcing platform. 
Each query requires annotations for 31 passages, and there are a total of 491,007 queries, leading to a total human annotation cost of \$1,369,910. 
We utilize cloud computing resources, where the cost of using an A800 80GB GPU is assumed to be \$0.8 per hour\footnote{\url{https://vast.ai/pricing/gpu/A800-PCIE}}. 
Our utility-focused annotation process requires a total of 53 hours on an 8 $\times$ A800 GPU machine using the Qwen-2.5-32B-Int8, resulting in a GPU computing cost of \$339. 
For the REPLUG method, the annotation process takes 70 hours, costing \$448 in GPU computing. 
However, REPLUG requires human-annotated answers for each query, bringing the total to \$44,639. 
More details are provided in Table~\ref{tab:cost}. 
Although human annotation achieves superior performance on the in-domain dataset, the cost of such annotation is substantial. 
In contrast, the utility-focused annotation offers the lowest annotation cost, with performance second only to that of human annotation.

\begin{table}[htbp]
\centering
\small
\renewcommand{\arraystretch}{0.8}
\setlength\tabcolsep{1pt}
    \begin{tabular}{l c ccc}
    \toprule
    Annotation	& Cost(\$) & Time(h) & MRR@10 & R@1000 \\
    \midrule
    Human &	1,369,910  & - & 38.6 &	98.6 \\
    REPLUG	& 44,639  & 70+ & 33.8 &  94.7\\
    \US	& 339 & 53& 35.3  &  97.7 \\
    \US (CL 20\%)	& 274,321 & - & 38.2  & 98.5 \\
    \bottomrule
\end{tabular}
\caption{Retrieval performance (\%) of different annotations on MS MARCO Dev and corresponding annotation cost. ``R@k'' means ``Recall@k''.}
\label{tab:cost}
\end{table}


\section{Prompts for Annotation via LLMs} 
\label{app:prompts}
\begin{figure*}[htbp]
    \centering
    \small
    \includegraphics[width=\linewidth]{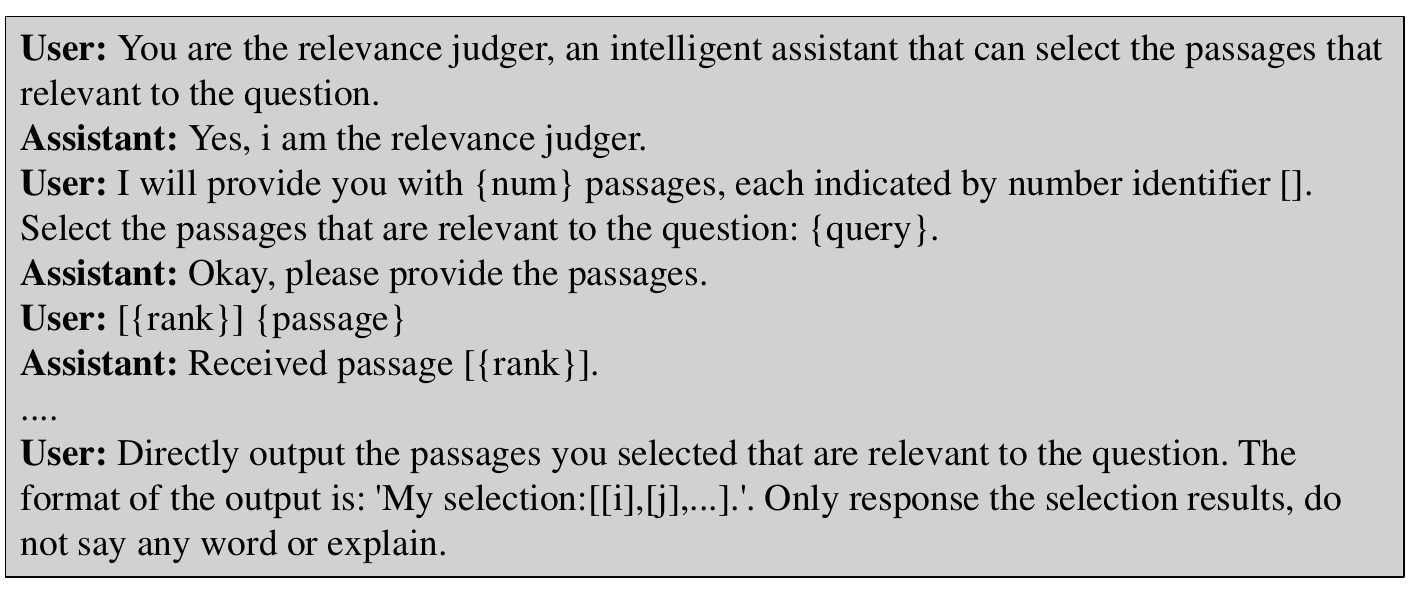}
    \caption{Relevance-based selection prompt for LLMs.}
    \label{fig:prompt-relevance-selection}
\end{figure*}

\begin{figure*}[htbp]
    \centering
    \small
    \includegraphics[width=\linewidth]{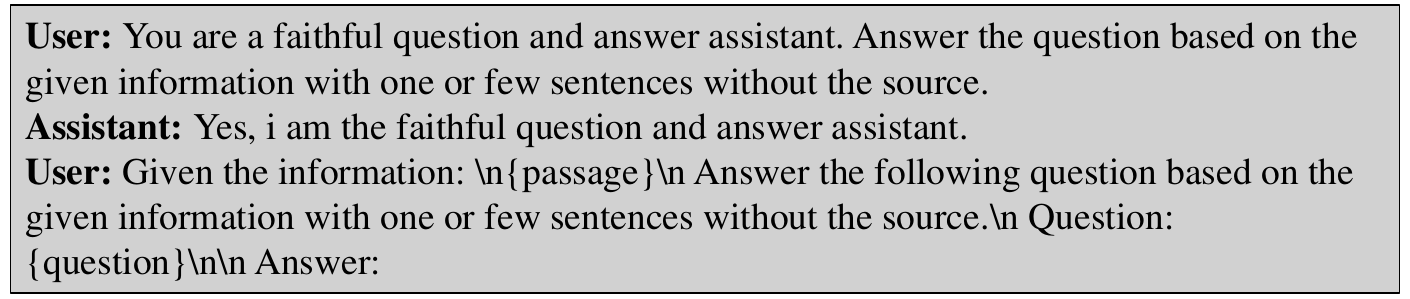}
    \caption{Pseudo-answer generation prompt for LLMs.}
    \label{fig:prompt-pseudo-answer}
\end{figure*}

\begin{figure*}[htbp]
    \centering
    \small
    \includegraphics[width=\linewidth]{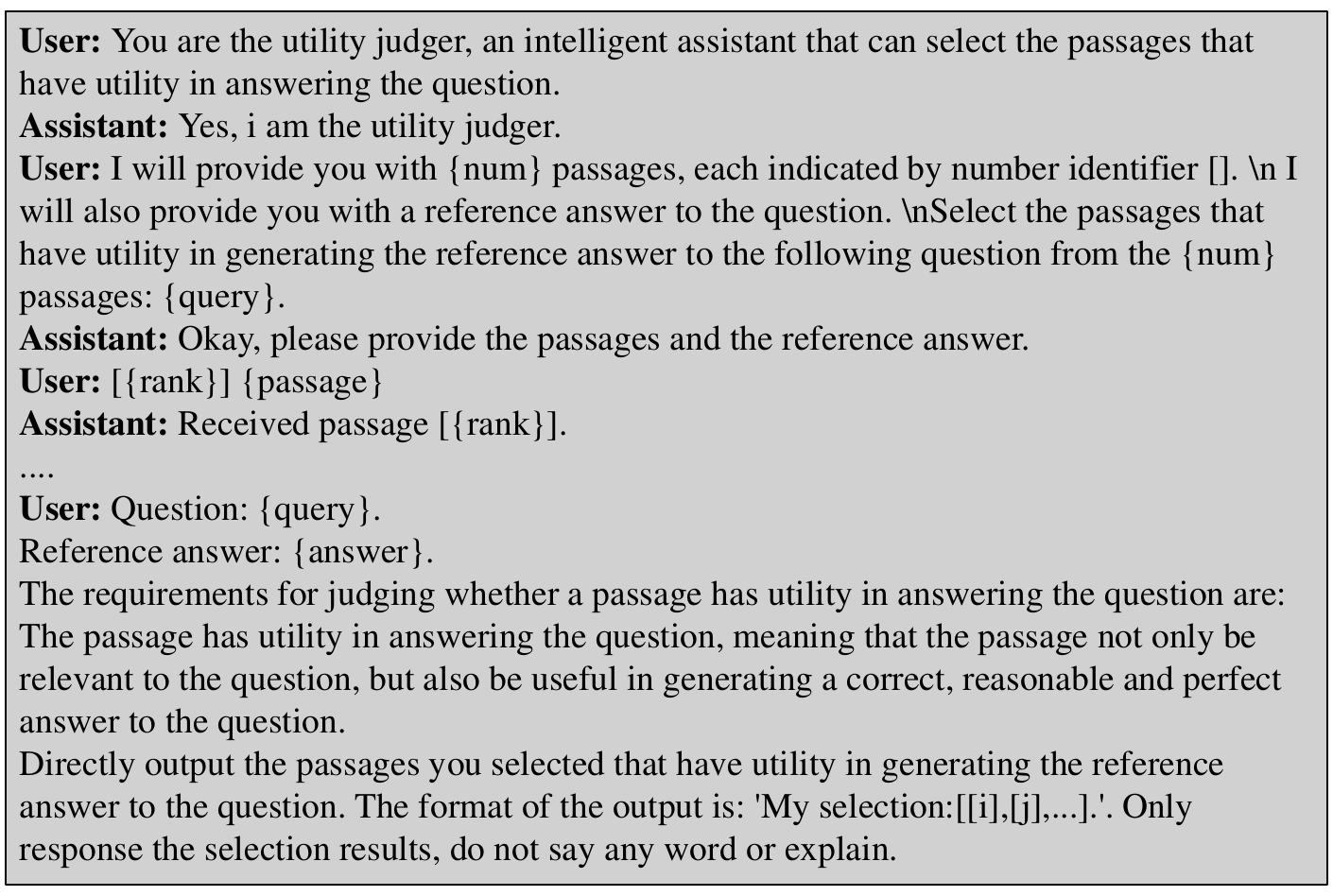}
    \caption{Utility-based selection prompt for LLMs.}
    \label{fig:prompt-utility-selection}
\end{figure*}

\begin{figure*}[htbp]
    \centering
    \small
    \includegraphics[width=\linewidth]{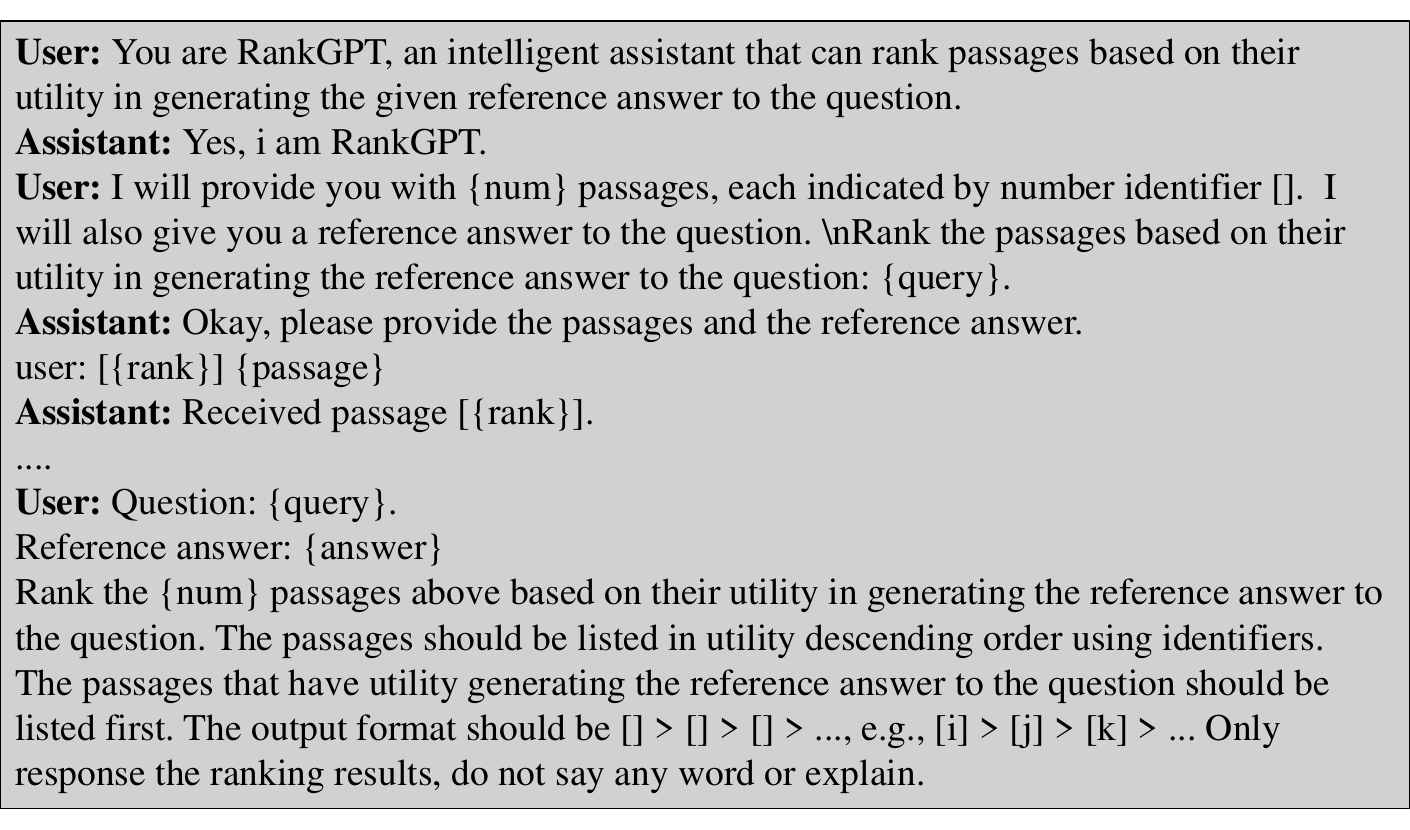}
    \caption{Utility-based ranking prompt for LLMs.}
    \label{fig:prompt-utility-ranking}
\end{figure*}

Relevance-based selection, pseudo-answer generation, utility-based selection, and utility-based ranking prompts are shown in Figure \ref{fig:prompt-relevance-selection}, Figure \ref{fig:prompt-pseudo-answer}, Figure \ref{fig:prompt-utility-selection}, and Figure \ref{fig:prompt-utility-ranking}, respectively.

\balance

\end{document}